\documentstyle[12pt,epsfig]{article}
\newcommand{\n}{\hspace*{-2.5mm}}
\newcommand{\simgt}{\rlap{\lower 3.5 pt \hbox{$\mathchar \sim$}} \raise 1pt
 \hbox {$>$}}
\newcommand{\simlt}{\rlap{\lower 3.5 pt \hbox{$\mathchar \sim$}} \raise 1pt
 \hbox {$<$}}

\catcode`@=11
\newcount\@tempcntc
\def\@citex[#1]#2{\if@filesw\immediate\write\@auxout{\string\citation{#2}}\fi
  \@tempcnta\z@\@tempcntb\m@ne\def\@citea{}\@cite{\@for\@citeb:=#2\do
    {\@ifundefined
       {b@\@citeb}{\@citeo\@tempcntb\m@ne\@citea\def\@citea{,}{\bf ?}\@warning
       {Citation `\@citeb' on page \thepage \space undefined}}%
    {\setbox\z@\hbox{\global\@tempcntc0\csname b@\@citeb\endcsname\relax}%
     \ifnum\@tempcntc=\z@ \@citeo\@tempcntb\m@ne
       \@citea\def\@citea{,}\hbox{\csname b@\@citeb\endcsname}%
     \else
      \advance\@tempcntb\@ne
      \ifnum\@tempcntb=\@tempcntc
      \else\advance\@tempcntb\m@ne\@citeo
      \@tempcnta\@tempcntc\@tempcntb\@tempcntc\fi\fi}}\@citeo}{#1}}
\def\@citeo{\ifnum\@tempcnta>\@tempcntb\else\@citea\def\@citea{,}%
  \ifnum\@tempcnta=\@tempcntb\the\@tempcnta\else
   {\advance\@tempcnta\@ne\ifnum\@tempcnta=\@tempcntb \else \def\@citea{--}\fi
    \advance\@tempcnta\m@ne\the\@tempcnta\@citea\the\@tempcntb}\fi\fi}
\catcode`@=12
\begin{document}
\title{\vskip-3cm{\baselineskip14pt
\centerline{\normalsize DESY~96--210\hfill ISSN~0418--9833}
\centerline{\normalsize CERN--TH/96--274\hfill}
\centerline{\normalsize MPI/PhT/96--103\hfill}
\centerline{\normalsize hep--ph/9610267\hfill}
\centerline{\normalsize October 1996\hfill}}
\vskip1.5cm
Large-$p_T$ Photoproduction of $D^{*\pm}$ Mesons in $ep$ Collisions}
\author{B.A. Kniehl$^1$, G. Kramer$^2$ and M. Spira$^3$\\
$^1$ Max-Planck-Institut f\"ur Physik (Werner-Heisenberg-Institut),\\
F\"ohringer Ring 6, 80805 Munich, Germany\\
$^2$ II. Institut f\"ur Theoretische Physik\thanks{Supported
by Bundesministerium f\"ur Forschung und Technologie, Bonn, Germany,
under Contract 05~7~HH~92P~(5),
and by EU Program {\it Human Capital and Mobility} through Network
{\it Physics at High Energy Colliders} under Contract
CHRX--CT93--0357 (DG12 COMA).},
Universit\"at Hamburg,\\
Luruper Chaussee 149, 22761 Hamburg, Germany\\
$^3$ Theoretical Physics Division, CERN,\\
1211 Geneva 23, Switzerland}
\date{}
\maketitle
\begin{abstract}
The cross section for the inclusive photoproduction of large-$p_T$ $D^{*\pm}$
mesons is calculated at next-to-leading order, adopting different approaches to 
describe the fragmentation of charm quarks into $D^{*\pm}$ mesons.
We treat the charm quark according to the massless factorization scheme, where
it is assumed to be one of the active flavours inside the proton and the
photon.
We present inclusive single-particle distributions in transverse momentum and
rapidity, including the contributions due to both direct and resolved photons.
We compare and assess the various implementations of fragmentation.
We argue that, in the high-$p_T$ regime, a particularly realistic description
can be obtained by convoluting the Altarelli-Parisi-evolved fragmentation
functions of Peterson et al.\ with the hard-scattering cross sections of 
massless partons where the factorization of the collinear singularities 
associated with final-state charm quarks is converted to the massive-charm
scheme.
The predictions thus obtained agree well with recent experimental
data by the H1 and ZEUS Collaborations at DESY HERA.

\end{abstract}
\newpage

\section{Introduction}

Heavy-quark production in the $ep$ colliding-beam experiments with DESY HERA
offers a novel way of testing perturbative quantum chromodynamics (QCD).
First results for the charm-quark photoproduction cross section
$\sigma(\gamma p\to c\bar c+X)$ were presented
by ZEUS \cite{der} and H1 \cite{aid}, and compared with a next-to-leading-order
(NLO) calculation \cite{fri}.
In this calculation, the massive-charm scheme has
been adopted, in which the charm-quark mass $m\gg\Lambda_{QCD}$ acts as a
cutoff and
sets the scale for the perturbative calculation. The cross section factorizes 
into a partonic hard-scattering cross section multiplied by light-quark
and gluon densities \cite{col}. In this factorization approach, the only quarks
inside the proton and the photon are the light ones.
Thus, in the massive-charm scheme, the number of active flavours in the initial
state is $n_f=3$, while the massive charm quark appears only in the final state.
For the prediction of the total charm photoproduction cross section, for which
experimental results from HERA were presented in \cite{der,aid},
this is the only possibility. Actually, $m\approx1.5$~GeV is not very large
compared to $\Lambda_{QCD}$, so
that the validity of this approach is not obvious.

With the advent of new measurements, by H1 \cite{aid} and ZEUS \cite{kar}, of
the differential cross section $d^2\sigma/dy\,dp_T$ of inclusive $D^{*\pm}$
production, where $y$ and $p_T$ are the rapidity and transverse momentum of the
$D^{*\pm}$ mesons, respectively, we have the possibility
to test the theory in a different regime of scales. The experimental
differential cross sections extend up to $p_T=12$~GeV, so that,
in contrast to the total-cross-section calculations,
$p_T$ rather than $m$ should be considered as the large scale.
Then, in NLO, terms proportional to $\alpha_s\ln(p_T^2/m^2)$ arise from
collinear gluon emission by charm quarks or from almost collinear branching
of gluons or photons into charm-anticharm pairs.
For large enough $p_T$, these terms are bound to spoil
the convergence of the perturbative series and cause large scale
dependences of the NLO result at $p_T\gg m$. The
proper procedure in the regime $p_T\gg m$ is to absorb the
terms proportional to $\alpha_s\ln(p_T^2/m^2)$ into the charm
distribution functions of the incoming photon and proton and into
the fragmentation functions (FF's) of charm quarks into charmed hadrons.
Of course, to perform this absorption, one needs information
on the charm contribution in the parton density functions (PDF's) and FF's.

An alternative way of making reliable predictions at
large $p_T$ is to treat the charm quarks as massless
partons. The collinear singularities corresponding to the
$\alpha_s\ln(p_T^2/m^2)$ terms of the massive-charm scheme
are then absorbed into the charm-quark PDF's and FF's
in the same way as for the lighter $u$, $d$ and $s$ quarks.
This massless approach was proposed in \cite{nas} and first
applied to the production of large-$p_T$
hadrons containing bottom quarks in $p\bar p$ collisions
by Lampe \cite{lam} and by Cacciari and Greco \cite{cac}. Subsequently, it
was employed in two independent studies of charm-quark photoproduction
\cite{kni,mca}. In these two investigations, the
$y$ and $p_T$ distributions in the massless and massive approaches were
compared with each other, making different assumptions concerning the
initial state as well as the FF's in the massless approach.
In \cite{kni}, low-$Q^2$ electroproduction was considered, while
\cite{mca} was concerned with photoproduction with fixed photon energy.
In \cite{kni}, the FF of the charm quark into charmed
hadrons was approximated by $\delta(1-z)$, where $z=p_D/p_c$ is the
scaled momentum of the charmed hadron $D$, whereas the authors of
\cite{mca} employed the perturbative FF's (PFF's) of \cite{mel},
which they evolved from the starting scale $m$ to the appropriate higher
scales of order $p_T$ according to the usual Altarelli-Parisi evolution
equations.

It is the purpose of this work to remove the restriction
to a scale-independent $\delta$-type FF
of the charm quark made in our previous study with Kr\"amer
\cite{kni} and to adopt more realistic descriptions of charm-quark
fragmentation, including evolution to higher scales.
Having extensively compared the massless and massive approaches in 
\cite{kni}, we shall now focus our attention on the massless approach,
which is likely to be more reliable in the large-$p_T$ range, in which we are
primarily interested here. Specifically,
we shall consider the following models of charm-quark
fragmentation:
(i) $\delta$-function-type FF without evolution \cite{kni} for reference;
(ii) PFF \cite{mel} with and without evolution; and
(iii) Peterson fragmentation \cite{pet} with evolution.
Choice (iii) will be considered as the most
realistic one and will be used for comparisons with the
recent H1 \cite{aid} and ZEUS \cite{kar} data.

The outline of our work is as follows. In Section~2, we shall
shortly describe the basic formalism of charm-quark fragmentation, discuss the
transition from massless to massive factorization of final-state collinear 
singularities, and specify our assumptions concerning
choices (i), (ii) and (iii).
In Section~3, we shall investigate numerically the differences between the
various implementations of charm-quark fragmentation discussed in Section~2.
There, we also confront our final predictions based on the Peterson FF
with the H1 and ZEUS data.
Our conclusions will be summarized in Section~4.

\section{Fragmentation function approach}

In this section, we describe the underlying assumptions
for the massless approach, making various choices for the
FF's of the (anti)charm quark into $D^{*\pm}$ mesons.
As is well known, two mechanisms contribute to the photoproduction
of charm quarks in $ep$ collisions: (i) In the direct
photoproduction mechanism, the photon couples directly to the quarks,
which, besides the massless $u$, $d$ and $s$ quarks, also include the
massless $c$ quark. In this case, no spectator particles
travel along the momentum direction of the photon. (ii)
In the resolved photoproduction mechanism,
the photon splits up into a flux of $u$, $d$, $s$, $c$ quarks and
gluons, which then interact with the partons coming
from the proton leading to the production of charm quarks at large $p_T$.
The contributing parton-level processes are the same as in the case of
charm-quark production in hadron-hadron collisions.
The charm quark is accompanied by a spectator jet in the photon direction.
Therefore, the $\gamma p$ cross
section depends not only on the PDF's
of the proton but also on those of the photon.
The main difference relative to the usually considered massive-charm scheme is
that the charm quark also contributes via the PDF's of
the proton and photon, i.e.\ charm is already
an active flavour in the initial state. In other words,
there is no essential difference between the treatment of the
truly light quarks $u$, $d$, $s$ and the charm quark in the initial
state. This approach is justified if a large scale is
governing the production process.
In our case, this is the $p_T$ of the $D^{*\pm}$ mesons, with $p_T\gg m$.
In this region, non-singular mass terms are suppressed in the
cross sections by powers of $m/p_T$. The important mass terms
appear if the virtuality of the charm quark is small.
This happens in the initial state if the charm quark is
emitted from the proton or photon, and in the final state
if the partons emit a $c\bar c$ pair. These two contributions
lead to the $\alpha_s\ln(p_T^2/m^2)$ terms in the massive scheme.
In our massless approach, they are summed into the scale-dependent
PDF's and FF's of charm quarks into $D^{*\pm}$ mesons.

Thus, when we proceed to NLO, the following steps are
taken: (i) The hard-scattering cross section for the direct-
and resolved-photon processes are calculated in the massless approximation
with $n_f=4$ active flavours. The collinear singularities
are subtracted according to the $\overline{\mbox{MS}}$ scheme. Since the
charm quark is taken to be massless, the singularities from its
splittings are subtracted as well.

(ii) The charm quark is accommodated
in the PDF's of the proton and photon as a light flavour.
The finite mass of the charm quark is taken into account by
including it in the evolution in such a way that
its PDF's are only non-vanishing above a scale set by its mass.

(iii) The FF's characterize the hadronization of the massless
partons, including the charm quark, into mesonic or baryonic states
containing the charm quark. In our case, these are the $D^{*\pm}$ mesons.
Similarly to the fragmentation into light mesons, these FF's are
basically non-perturbative input and must be determined by experiment,
for example from the cross section of inclusive $D^{*\pm}$ production
in $e^+e^-$ annihilation \cite{laf}. This information
determines the FF's at some starting scale, $\mu_0$.

An alternative that exploits the fact that the charm mass satisfies
$m\gg\Lambda_{QCD}$ is the calculation of universal starting
conditions for the FF's within perturbative QCD
at a scale $\mu_0$ of order $m$.
The FF's thus obtained are the PFF's mentioned above.
In \cite{mel},
these starting conditions were calculated at NLO in the
$\overline{\mbox{MS}}$ scheme. They read
\begin{eqnarray}
\label{eqpff}
D_c^D(x,\mu_0)&\n=\n&D_{\bar c}^{\bar D}(x,\mu_0)
=\delta(1-x)+\frac{\alpha_s(\mu_0)}{2\pi}C_F
\left\{\frac{1+x^2}{1-x}\left[\ln\frac{\mu_0^2}{m^2}-2\ln(1-x)-1\right]
\right\}_+,
\nonumber\\
D_g^D(x,\mu_0)&\n=\n&D_g^{\bar D}(x,\mu_0)
=\frac{\alpha_s(\mu_0)}{2\pi}T_f[x^2+(1-x)^2]\ln\frac{\mu_0^2}{m^2},
\nonumber \\
D_{q,\bar q,\bar c}^D(x,\mu_0)&\n=\n&D_{q,\bar q,c}^{\bar D}(x,\mu_0)
=0.
\end{eqnarray}
Here, $D_i^D$ ($D_i^{\bar D}$) refers to the fragmentation of parton $i$
into a hadron $D$ ($\bar D$) containing a $c$ ($\bar c$) quark, $C_F=4/3$,
$T_f=1/2$, and $q=u,d,s$.
The normalization of these FF's is such that $D$ stands for the sum of all
hadrons with a $c$ quark.
To obtain the FF into a specific charmed meson, for example the $D^{*+}$,
we must include the branching ratio $B(c\to D^{*+})$.

(iv) For the higher scales $\mu>\mu_0$, the PDF's, the non-perturbative
FF's and the PFF's are evolved in NLO up to the chosen
factorization scale (which we take to be of order $p_T$)
via the Altarelli-Parisi equations and convoluted
with the NLO hard-scattering cross sections. The
charm-quark mass $m$ then only enters in terms of the starting
conditions of the PDF's and FF's.

As already mentioned, the large logarithmic terms
proportional to $\ln(p_T^2/m^2)$, which appear in the massive scheme,
are resummed in this approach.
To be specific, they are effectively split into two parts.
One part proportional to $\ln(p_T^2/\mu^2)$,
where $\mu$ is some generic factorization scale, appears in
the hard-scattering cross sections having no dependence on $m$.
This part may be kept small by choosing $\mu\approx p_T$. The other part,
proportional to $\ln(\mu^2/m^2)$, is absorbed into the PDF's and FF's.
The logarithm $\ln(\mu^2/\mu_0^2)$, which is large if
$\mu_0\approx m$ and $\mu\approx p_T$, is incorporated
via the evolution equations and therefore resummed.
The residual term proportional to $\ln(\mu_0^2/m^2)$,
connected with the starting conditions in (\ref{eqpff}), is treated
with fixed-order perturbation theory in the case of the PFF's,
or is part of the non-perturbative input in the case of general FF's.

The PFF's given in terms of the starting conditions (\ref{eqpff}) suggest
some simplifications which we consider as reasonable approximations for the
PFF's.
The dominant term in (\ref{eqpff}) is the $\delta$ function expressing the
fact that, due to $m\gg\Lambda_{QCD}$, the charm quark transforms into the
charmed hadron having the same momentum, any binding effect of the light quark
of order $\Lambda_{QCD}$ being neglected.
The approximation $D_c^D(x,\mu)=\delta(1-x)$, denoted $A$ for later use,
was considered in our earlier work \cite{kni}. This approximation
neglects the important $\ln(\mu^2/\mu_0^2)$ terms induced in the FF's
through the evolution.
Unless otherwise stated, we shall choose $\mu_0=2m$ and 
$\mu=2m_T$, where
$m_T=\sqrt{m^2+p_T^2}$ is the transverse mass of the produced hadron.

Approximation $B$ consists of keeping all terms in (\ref{eqpff}), but with
the replacement $\mu_0=\mu$.
This approximation includes the large $\ln(\mu^2/m^2)$ terms to fixed order in
$\alpha_s$, i.e.\ not in the resummed form because the FF's are not evolved up
to the scale $\mu$ via the evolution equations. Except
for the fact that the proton and photon PDF's still have four active flavours,
this approximation is, from the conceptual point of view, most closely
related to the massive scheme \cite{fri,kni}, in the sense that the important
mass terms of the massive calculation, namely the logarithmic ones related to
the collinear would-be singularities, are properly included, while the 
non-singular power terms in $m/p_T$ are omitted.
We stress that the latter are indispensible if $p_T$ is of order $m$, and must
be included if total cross sections are to be calculated;
approximation $B$ is then bound to break down.
However, our goal is to improve the theoretical description at $p_T\gg m$.
Since, in case $B$, the FF's are not evolved to higher
scales according to the evolution equations, the terms proportional to
$\alpha_s$ in (\ref{eqpff}) may be shifted to the hard-scattering
cross section, which we actually choose to do.
In fact, case $B$ may be viewed as a change of factorization scheme, in the 
sense that the factorization of the final-state collinear singularities
associated with the charm quark is adjusted so as to match the
finite-$m$ calculation.
This is an equivalent interpretation of the matching procedure
proposed in \cite{mel} between the
massless-charm calculation in connection with the PFF's and the massive-charm
calculation without FF's.
Specifically, we substitute in the hard-scattering cross sections
$\alpha_s(\mu)P_{ci}^{(0,T)}(x)\ln(s/\mu^2)\to
\alpha_s(\mu)P_{ci}^{(0,T)}(x)\ln(s/\mu^2)
-\alpha_s(\mu_0)d_{ci}(x)$,
where $P_{ci}^{(0,T)}(x)$ are the timelike LO $i\to c$ splitting functions
\cite{alt} and $d_{ci}(x)$ may be gleaned from (\ref{eqpff}):
\begin{eqnarray}
\label{eqdij}
d_{cc}(x)&\n=\n&
-C_F\left\{\frac{1+x^2}{1-x}\left[\ln\frac{\mu_0^2}{m^2}-2\ln(1-x)-1\right]
\right\}_+
\nonumber\\
&\n=\n&-P_{qq}^{(0,T)}(x)\ln\frac{\mu_0^2}{m^2}
+C_F\left\{-2\delta(1-x)+2\left(\frac{1}{1-x}\right)_+
+4\left[\frac{\ln(1-x)}{1-x}\right]_+\right.
\nonumber\\
&\n\n&{}-\left.\vphantom{\left(\frac{1}{1-x}\right)_+}
(1+x)\left[1+2\ln(1-x)\right]\right\},
\nonumber\\
d_{cg}(x)&\n=\n&-T_f[x^2+(1-x)^2]\ln\frac{\mu_0^2}{m^2}
\nonumber\\
&\n=\n&-P_{qg}^{(0,T)}(x)\ln\frac{\mu_0^2}{m^2},
\nonumber\\
d_{cq}(x)&\n=\n&d_{c\bar q}(x)=d_{c\bar c}(x)=0,
\end{eqnarray}
with $q=u,d,s$.
The $d_{\bar ci}$ functions emerge by charge conjugation.
Since the PFF's of (\ref{eqpff}) refer to the $\overline{\mbox{MS}}$
factorization scheme for the collinear singularities associated with outgoing
massless-charm lines, so do the $d_{ij}$ functions of (\ref{eqdij}).
In case $B$, $\ln(\mu_0^2/m^2)\approx\ln(p_T^2/m^2)$ is large.
Leaving aside the terms in (\ref{eqdij}) that are not enhanced by this 
logarithm, case $B$ obviously amounts to choosing the final-state
factorization scale in the massless-charm hard-scattering cross sections
to be $\mu=m$ and omitting the $d_{ij}$ functions, i.e.\ to choosing the
final-state factorization scale in case $A$ to be $\mu=m$. 
In our implementation of case $B$, we have $D_c^D(x,\mu)=\delta(1-x)$ for all
factorization scales $\mu$, just as in case $A$.

Obviously, case $B$ may be refined by resumming the large logarithms
$\ln(p_T^2/m^2)$ via evolution.
To this end, in case $C$, we take $D_c^D(x,\mu_0)=\delta(1-x)$ as the input
distribution at the starting scale $\mu_0$ for the NLO evolution up to $\mu$.
At the same time, we keep the $d_{ij}$ functions, with the large logarithm
stripped off, i.e.\ with $\mu_0$ now of order $m$ rather than $p_T$,
in the hard-scattering cross sections.

An alternative refinement of case $B$ may be obtained by taking the full PFF
distributions of (\ref{eqpff}) as input for the evolution
up to scale $\mu$, and convoluting the outcome with the massless-charm
hard-scattering cross sections (without $d_{ij}$ functions).
This approach, which we denote by $D$, corresponds to the PFF approach
advocated in \cite{mel} for inclusive heavy-quark production via $e^+e^-$
annihilation.
It was applied to inclusive bottom-quark production in $p\bar p$ collisions
in \cite{cac} and to inclusive charm-quark production in $\gamma p$ and
$\gamma\gamma$ collisions in \cite{mca,cgk}, respectively.

However, it is not at all clear that the PFF's
give the correct description for the fragmentation
of the charm quark into charmed hadrons, since the charm
quark is only moderately heavy. Although the average value
$\langle x\rangle$ near $\mu=\mu_0$ is not precisely known, it is bound
to be $\langle x\rangle<1$. Most measurements of $\langle x\rangle$ are at
larger scales. In the case of $c\to D^{*+}$, we have
$\langle x\rangle=0.495{+0.010\atop-0.011}\mbox{(stat.)}\pm0.007\mbox{(sys.)}$
from ALEPH \cite{bus} and
$\langle x\rangle=0.515{+0.008\atop-0.005}\mbox{(stat.)}\pm0.010\mbox{(sys.)}$
from OPAL \cite{ake} at $\mu=91.2$~GeV, and $\langle x\rangle=0.73\pm0.05$
at $\mu=10$~GeV \cite{alb}, indicating that $\langle x\rangle<1$ at
$\mu=\mu_0$ as well.
Therefore, the distribution by Peterson et al.\ \cite{pet} is usually
considered to be a better approximation for the FF at
the starting scale $\mu_0$. It has the form
\begin{equation}
\label{eqpssz}
D_c^D(x,\mu_0)={\cal N}\frac{x(1-x)^2}{\left[(1-x)^2+\epsilon x\right]^2}.
\end{equation}
Apart from the normalization $\cal N$, it depends just on
the parameter $\epsilon$, which is related to
$\langle x\rangle$.
Note that (\ref{eqpssz}) turns into a $\delta$ function in the limit 
$\epsilon\to0$.
We choose $\epsilon=0.06$, which is the central value
obtained in \cite{chr} for charm quarks.
In case $E$, we adopt (\ref{eqpssz}) with
\begin{equation}
{\cal N}^{-1} =
\frac{(\epsilon^2-6\epsilon+4)}{(4-\epsilon) \sqrt{4\epsilon-\epsilon^2}}
\left\{
\arctan\frac{\epsilon}{\sqrt{4\epsilon-\epsilon^2}}
+ \arctan\frac{2-\epsilon}{\sqrt{4\epsilon-\epsilon^2}} \right\}
+ \frac{1}{2} \ln \epsilon + \frac{1}{4-\epsilon},
\end{equation}
so that $\int_0^1dx\,D_c^D(x,\mu_0)=1$.
For $\epsilon=0.06$, one has ${\cal N}\approx0.625$.
This simplifies the comparison with the fragmentation models $A$--$D$.
In Section~3.4, where we compare our predictions with the HERA data,
we must adjust the normalization of our calculation by including the
measured branching fraction $B(c\to D^{*+})$ \cite{ake}.

The Peterson FF for $c\to D^{*+}$ fragmentation of (\ref{eqpssz}) has also
been used in \cite{sfr} for calculating the production of $D^{*\pm}$
mesons in $\gamma p$ collisions based on the purely massive scheme.
In our approach, we need the evolved FF's
at the scale $\mu$, with (\ref{eqpssz}) describing the input at
the starting scale $\mu_0$. We solve the evolution equations
using moments in Mellin space. For the reader's convenience,
the moments of (\ref{eqpssz}) are listed in
Appendix~A. They are expressed in terms of hypergeometric functions
of complex arguments, which are conveniently calculated from appropriate series
expansions, also written down in Appendix~A. To account for
the fact that the Peterson FF describes the transition from
a massive charm quark into a physical $D^{*+}$ meson,
whereas, for reasons explained in the Introduction, we work with
hard-scattering cross sections where the charm quark is treated as a massless
flavour, we incorporate the $d_{ij}$ functions of (\ref{eqdij}) in our
analysis just as we do in case $C$.
This means that case $E$ emerges from case $C$ by replacing the
$\delta$-function-type input distribution at scale $\mu_0$ with
the Peterson formula of (\ref{eqpssz}).

An alternative approach would be to include the terms relating the massless
and massive schemes in the evolution of the FF as we do in version $D$, where
we evolve the full PFF's of (\ref{eqpff}).
In other words, one would convolute the PFF's with the Peterson
formula~(\ref{eqpssz}) at the scale $\mu_0$, evolve these modified initial
distributions at NLO up to the scale $\mu$ and employ the FF's thus obtained
in the usual massless-charm $\overline{\mbox{MS}}$ analysis (without $d_{ij}$
functions).
In the following, this possible alternative for scheme $E$ will be called
scheme $F$.
Both of these schemes properly resum the large logarithmic terms of the type
$\ln(\mu^2/\mu_0^2)$ which emerge from (\ref{eqpff}) as $\mu_0$ is driven
up to $\mu$.
In scheme $E$, just the latter logarithms get resummed, while the
non-logarithmic terms of (\ref{eqpff}) are treated in fixed-order
perturbation theory.
By contrast, in scheme $F$, also these non-logarithmic terms are included in 
the evolution.
Scheme $E$ is minimal in the sense that is allows us to include in the
evolution just the long-distance part of the $c\to D^{*+}$ fragmentation
process, which is encoded in the Peterson ansatz~(\ref{eqpssz}).
We shall adopt scheme $E$ whenever we confront theoretical predictions with
HERA data.
In the limit $\epsilon\to0$, cases $E$ and $F$ coincide with cases $C$ and
$D$, respectively.
For a given value of $\epsilon$, one expects the difference between schemes
$E$ and $F$ to be similar to the one between schemes $C$ and $D$.
Since schemes $C$ and $D$ do not yield physical observables, their difference
is phenomenologically inconsequential.
By the same token, it is unreasonable to expect that a cross section evaluated
in schemes $C$ or $D$ should coincide with the corresponding charm-quark
production cross section obtained in the all-massive scheme.
In the case of $D^{*\pm}$ photoproduction, this is even more apparent, since 
the cross sections of schemes $C$ and $D$ exhibit a significant sensitivity to
the charm content of the resolved photon (intrinsic charm), which does not 
enter the all-massive calculation at all.
On the other hand, the difference between schemes $E$ and $F$ is to be
compensated by a corresponding shift in the $\epsilon$ parameter of
(\ref{eqpssz}), i.e.\ $\epsilon$ is scheme dependent at NLO.
For instance, if one wishes to use the value of $\epsilon$ determined through 
a NLO fit to $e^+e^-$ data as input for a NLO prediction of photoproduction,
one has to make sure that both analyses are carried out in the same scheme.
Only then, one may obtain a coherent description of high-energy $e^+e^-$ data
and high-$p_T$ $ep$ data.
We emphasize that, on the basis of an all-massive calculation including
fragmentation, this cannot be achieved in any meaningful way, due to the
presence of sizeable logarithms at fixed order.
A consistent procedure within the all-massive framework would be to determine
$\epsilon$ from a NLO fit to low-energy $e^+e^-$ data on inclusive $D^{*\pm}$
production, which has not been done so far.

A second alternative to case $E$ has been adopted in \cite{rol} to 
describe charmed-meson production at the Tevatron.
In \cite{rol}, the evolved PFF's are convoluted with some unevolved
$x$ distribution of the form $N(1-x)^\alpha x^\beta$.
By contrast, we assume that the non-perturbative $x$ distribution of the
$D^{*+}$ mesons, where $x$ is the longitudinal-momentum fraction w.r.t.\ the
quasi-massive parent charm quarks, is subject to evolution.

The calculation of the hard-scattering cross sections proceeds along the lines
of our previous works \cite{kni,bor} on the basis of the direct- and 
resolved-photon hard-scattering cross section obtained in \cite{aur} and
\cite{ave}, respectively.

\section{Results}

This section consists of four parts.
First, we specify our assumptions concerning the proton and photon PDF's
as well as the equivalent photon approximation.
Second, we describe the various degrees of refinement in the treatment of
fragmentation.
Third, we discuss various theoretical uncertainties related to the proton and 
photon PDF's and the dependence on scales and parameters intrinsic to the
FF's.
Finally, in the fourth part, we present our predictions for the inclusive
$D^{*\pm}$ photoproduction under realistic kinematic conditions corresponding
to the H1 and ZEUS experiments.

\subsection{Input information}

For the calculation of the cross section $d^2\sigma/dy_{lab}\,dp_T$, we adopt
the present standard HERA conditions, where $E_p = 820$~GeV protons collide
with $E_e = 27.5$~GeV positrons in the laboratory frame.
We sum over $D^{*+}$ and $D^{*-}$ mesons.
We take $y_{lab}$, which refers to the laboratory frame, to be positive in the
proton flight direction.
The quasi-real-photon spectrum is described in the Weizs\"acker-Williams
approximation with the formula
\begin{equation}
\label{eqwwa}
f_{\gamma/e}(x) = \frac{\alpha}{2\pi}\left[\frac{1+(1-x)^2}{x}
\ln\frac{Q_{max}^2}{Q_{min}^2}
+2m_e^2x\left(\frac{1}{Q_{max}^2}-
              \frac{1}{Q_{min}^2}\right)\right],
\end{equation}
where $x = E_\gamma/E_e$, $Q_{min}^2 = m_e^2x^2/(1-x)$,
$\alpha$ is Sommerfeld's fine-structure constant and $m_e$ is the electron
mass.
In the case of ZEUS, where the final-state positron is not detected
(ZEUS untagged), one has $Q_{max}^2 = 4$~GeV$^2$ and
$0.147 < x < 0.869$, which corresponds to $\gamma p$ centre-of-mass (CM)
energies of 115~GeV~$< W <$~280~GeV \cite{kar}.
We shall use these conditions for the evaluation of 
the cross sections in the next two subsections.
Besides the untagged case, the H1 Collaboration has also presented data where
the positron is tagged.
The corresponding parameters will be specified later.
Our default sets for the proton and photon PDF's are
CTEQ4M \cite{lai}, with
$\Lambda_{\overline{\mbox{\scriptsize MS}}}^{(4)}=296$~MeV, and GRV \cite{grv}.
We work at NLO in the $\overline{\mbox{MS}}$ scheme with $n_f=4$ flavours.
We identify the factorization scales associated with the proton, photon and
final-state hadron, $M_p$, $M_\gamma$ and $M_h$, respectively,
and collectively denote them by $M_f$.
Our standard choice of renormalization and factorization scales is
$\mu=\xi m_T$ and $M_f=2\xi m_T$ with $\xi=1$.
If the FF's are evolved, we take the starting scale to be $\mu_0=2m$ with
$m=1.5$~GeV.
We calculate $\alpha_s(\mu)$ from the two-loop formula with
$\Lambda_{\overline{\mbox{\scriptsize MS}}}^{(4)}$ equal to the value used in
the proton PDF's.

\subsection{Refinements in fragmentation}

In our previous work \cite{kni}, we made the most simple ansatz for the 
fragmentation of charm quarks into charmed hadrons, namely
$D_c^D(x,\mu)=\delta(1-x)$ for all scales $\mu$.
In Section~2, this was referred to as case $A$.
In Figs.~\ref{fig1} and \ref{fig2}, we show the predictions for
$d\sigma/dp_T$ integrated over $-1.5<y_{lab}<1$ and
$d^2\sigma/dy_{lab}\,dp_T$ at $p_T=7$~GeV, respectively, according to this case
together with the refinements $B$--$F$ discussed in the same section,
adopting the ZEUS untagged conditions.
Cases $B$ and $C$ differ from case $A$ in that they take the evolution into
account.
In the rough approximation of case $B$, the evolution is incorporated by 
choosing $\mu_0=M_h$ in the definitions (\ref{eqdij}) of the $d_{ij}$
functions,
which correct for massive final-state factorization in the hard-scattering
cross sections.
This generates large logarithms of the form $\ln(M_h^2/m^2)$, which are
resummed in case $C$ by properly evolving the $\delta$-type FF's from the
starting scale $\mu_0=2m$ up to $M_h$.
In case $C$, the transition to the massive factorization is performed by
adding the $d_{ij}$ functions with $\mu_0=2m$.
In contrast to the case $B$, case $C$ includes also contributions from gluon 
fragmentation and, to a lesser extent, from light-quark fragmentation.
In case $D$, the hard-scattering cross sections in the massless
factorization scheme ($d_{ij}=0$) are convoluted with the evolved
PFF's of (\ref{eqpff}).
The effect of the $d_{ij}$ functions is then contained in the PFF's.

We see in Fig.~\ref{fig1} that, at low $p_T$, the latter three cases
($B$, $C$ and $D$) give very similar cross sections $d\sigma/dp_T$,
whereas there are appreciable differences at large $p_T$, especially between
cases $C$ and $D$.
Case $D$ exhibits the strongest fall-off.
Here not only the $\delta$ function, but also the finite, perturbative terms
of (\ref{eqpff}) are included in the evolution.
Case $B$, where evolution is accounted for only in the leading-logarithmic
approximation, has the smallest decrease of $d\sigma/dp_T$ with increasing
$p_T$.
Case $C$, where only the $\delta$ function is evolved,
lies between cases $B$ and $D$, as one would expect, but closer to case $B$.
We emphasize that the evolution of the FF's is performed up to NLO
(cases $C$ and $D$). We conclude from this study that the evolution
of the FF must be fully taken into account in order to obtain
realistic predictions for the $p_T$ distribution. The unevolved
$\delta$-function fragmentation considered in \cite{kni} (case $A$) is
unreliable for large $p_T$.

In \cite{kni}, case $A$ was compared with the massive NLO calculation,
adopting the previous ZEUS conditions, which are very similar to the present
ones used here.
The massive NLO result turned out to be roughly 45\% of the case-$A$ result.
On the other hand, from Fig.~\ref{fig1} we see that, at $p_T=3$~GeV (12~GeV),
the case-$B$ result amounts to 67\% (54\%) of the case-$A$ result.
In other words, at $p_T=3$~GeV (12~GeV), the massive NLO result is roughly
2/3 (5/6) of the case-$B$ result.
This may be partly attributed to the negative contribution from the power
terms in $m/p_T$, which are not included in case $B$.
The magnitude of these terms rapidly decreases with increasing $p_T$.
On the other hand, in contrast to the massive calculation, case $B$ receives
contributions from the charm quarks intrinsic to the proton and photon.
These contributions will be investigated in Figs.~\ref{fig4} and \ref{fig5},
respectively.
We shall see that the charm content of the photon is particularly important.

The rapidity distributions at $p_T=7$~GeV for the cases $A$, $B$, $C$ and $D$
are shown in Fig.~\ref{fig2}.
Here the pattern is somewhat different. In the $y_{lab}$
region where the cross section $d^2\sigma/dy_{lab}\,dp_T$ is maximal, i.e.\
$-1.5<y_{lab}<1$, the distributions of cases $A$, $B$, $C$ and $D$ have very
similar shapes.
The overall normalizations just differ according to the
hierarchy in Fig.~\ref{fig1}. For the larger rapidities $y_{lab}>2$, the cross
sections for cases $A$, $B$, $C$ and $D$ almost coincide, i.e.\ here the
evolution of the FF's has little effect.

In Figs.~\ref{fig1} and \ref{fig2}, we have also included the predictions
based on the Peterson FF evolved to larger scales in NLO, corresponding to
cases $E$ and $F$.
This FF is also normalized to unity, as all the other FF's, in order to
facilitate the comparison.
That is, in the limit $\epsilon\to0$, the results of cases $E$ and $F$ 
coincide with those of cases $C$ and $D$, respectively.
As anticipated in Section~2, for finite $\epsilon$, the relative difference
between cases $E$ and $F$ is very similar to that between cases $C$ and $D$.
In particular, the $p_T$ distributions for cases $E$ and $F$ in
Fig.~\ref{fig1} converge towards the low-$p_T$ end, where the non-logarithmic
terms of the PFF's as implemented in case $F$ are just mildly affected by the
evolution.
The main effect of the Peterson FF with finite $\epsilon$ is to reduce the
overall normalization of the respective calculations.

In the following investigations, we shall stick to case $E$ and employ the
NLO-evolved Peterson FF.
This description of charm fragmentation is more realistic than the PFF's
because the charm quark is only moderately heavy.
Compared with the massive-charm scheme, the massless-charm scheme shows
a completely different decomposition of the cross section
into direct- and resolved-photon contributions. This was analysed
in detail in our earlier work \cite{kni}. Whereas in the massive
approach direct photoproduction is dominant at large
$p_T$, in our massless scheme both contributions are of the
same order of magnitude \cite{kni}. This is demonstrated
in Fig.~\ref{fig3} for $d^2\sigma/dy_{lab}\,dp_T$ at $p_T=7$~GeV.
Both cross sections peak approximately at $y_{lab}=0$.
The resolved cross section has in addition a shoulder
at larger $y_{lab}$, which originates from the gluon part of
the photon PDF's. The peak at
$y_{lab}=0$ is due to the charm part, as may be inferred from Fig.~\ref{fig5}.
Of course, we must bear in mind that the decomposition of the photoproduction
cross section in direct- and resolved-photon contributions is ambiguous at 
NLO; it depends on the factorization scheme and scale, which we take to be the
$\overline{\mbox{MS}}$ scheme and $M_\gamma=2m_T$, respectively.
However, the sum of the two contributions is unambiguously defined.

\subsection{Theoretical uncertainties}

Before we compare our results with recent experimental
data from H1 and ZEUS, we investigate several theoretical
uncertainties, which might be relevant.
First, we consider the influence of the proton PDF's.
In Fig.~\ref{fig4}, the cross section $d^2\sigma/dy_{lab}\,dp_T$ at
$p_T=7$~GeV is plotted for our standard CTEQ4M \cite{lai} proton PDF's
and compared with the result obtained using the recent proton PDF set MRS(G)
by the Durham group \cite{mrs},
with $\Lambda_{\overline{\mbox{\scriptsize MS}}}^{(4)}=254$~MeV.
The cross sections for these two choices
are almost identical. From this we conclude that our
predictions are insignificantly influenced through our choice of the
proton PDF's. In Fig.~\ref{fig4}, we also show the influence
of the charm distribution inside the proton. If it is taken
away from the CTEQ4M PDF's, the cross section
hardly changes. Only for larger $y_{lab}$, where
the cross section is small, do we see the effect of the
charm content of the proton.

A similar study w.r.t.\ the photon PDF's is presented in Fig.~\ref{fig5}.
Here we compare $d^2\sigma/dy_{lab}\,dp_T$ at $p_T=7$~GeV,
evaluated with our standard GRV photon PDF's, with the calculation based on
the recently published new version of the photon PDF's
by Gordon and Storrow \cite{gor}. The cross sections
differ somewhat, by typically 10--20\%, over
the whole $y_{lab}$ range. This means that very accurate
data are needed if the differences between the GRV and
GS photon PDF's are to be disentangled.
In Fig.~\ref{fig5}, we have also plotted
the GRV prediction with the charm
distribution inside the photon put to zero. This
has a dramatic effect. For $y_{lab}<1$, the curves for the direct-photon
contribution and the sum of the direct and resolved contributions are almost
identical.
This shows that the resolved cross section is essentially made up by the charm
content of the photon.
Only for $y_{lab}>1$, where the cross section decreases with increasing
$y_{lab}$, can we see the effect of the other components of the photon, i.e.\
the gluon and the light quarks. Therefore
charmed-hadron photoproduction at large $p_T$ seems
to be an ideal place to learn specifically about the
charm content of the photon \cite{kni,dre},
which otherwise can only be studied in charmed-hadron production in
large-$Q^2$ $e\gamma$ scattering.

All finite-order perturbative calculations are plagued
by scale dependences. Although we may expect that
this dependence is reduced in NLO, it is not in general negligible, in
particular at moderate $p_T$. In Fig.~\ref{fig6}, we show the
cross section $d^2\sigma/dy_{lab}\,dp_T$ at $p_T=7$~GeV for our standard
PDF choice and scales $\mu=M_f/2=\xi m_T$ with $\xi=1/2$, 1 and 2.
In all these cases, we keep the constraint $\mu_0=2m$.
This scale variation produces changes of the cross section
of only $\pm10\%$, which indicates good perturbative stability.
The maximum change occurs where the cross section is maximal.
The main effect is actually due to the variation of the renormalization scale.
In fact, if we stick to $\mu=m_T$, but choose $M_f=m_T$ instead of $M_f=2m_T$,
we observe only a small difference in the cross section; compare the dotted
and full curves in Fig.~\ref{fig6}.

The cross section for the production of charmed hadrons depends on the choice
of FF's.
This was already demonstrated in Figs.~\ref{fig1} and \ref{fig2},
where we compared the cross sections for different assumptions concerning the
FF's.
Possibly, the choice $\epsilon = 0.06$ in the Peterson FF, which we have
adopted from \cite{chr}, might not be fully compatible with our procedure.
In particular, the value of $\epsilon$ must depend on the value of the initial
scale $\mu_0$.
Up to now, we have fixed $\mu_0=2m$ and $\epsilon =0.06$.
In Fig.~\ref{fig7}, we study how variations in $\mu_0$ and $\epsilon$ affect 
the cross section $d^2\sigma/dy_{lab}\,dp_T$ at $p_T=7$~GeV.
If we reduce $\mu_0$ to $\mu_0=m$ keeping $\epsilon =0.06$, the cross section
decreases, as may be seen by comparing the dotted curve with the full one.
This reduction can be compensated by simultaneously adjusting $\epsilon$ to a
smaller value.
The variation of the cross section with $\epsilon$ may also be inferred from
Fig.~\ref{fig7}, where $\epsilon = 0.04$, 0.06 and 0.08 are considered.
Decreasing $\epsilon$ leads to a larger cross section.
This may be understood by observing that case $C$ emerges from case $E$ in the
limit $\epsilon\to0$.
In order to obtain definite predictions, we must use the information on the
$\epsilon$ parameter in connection with a fixed choice for $\mu_0$ from other
$D^{*\pm}$ production experiments, e.g.\ from $e^+e^-\to D^{*\pm}+X$.
So far, the choice $\epsilon=0.06$ was motivated by Chrin's analysis
\cite{chr} of charmed-meson production in the PETRA energy range.
This analysis relied on fragmentation models.
Therefore it is unclear whether $\epsilon=0.06$ is the correct choice for the
Peterson FF at $\mu_0=2m$.
We shall come back to this point when we compare our calculation with the
experimental data from H1 and ZEUS.

\subsection{Comparison with H1 and ZEUS data}

In this section, we compare our NLO predictions for the cross section of
inclusive $D^{*\pm}$ photoproduction in $ep$ scattering at HERA with three
recent sets of data:
a) H1 tagged, b) H1 untagged \cite{aid} and c) ZEUS untagged \cite{kar}.
We sum over $D^{*+}$ and $D^{*-}$ mesons and adjust the normalization of the
Peterson FF by including the measured branching fraction
$B(c\to D^{*+})=B(\bar c\to D^{*-})=0.260$ \cite{ake}.
We adopt the respective experimental constraints on the equivalent photon 
spectrum of (\ref{eqwwa}):
a) $0.28<x<0.65$, $Q_{max}^2=0.01$~GeV$^2$,
b) $0.1<x<0.8$, $Q_{max}^2=4$~GeV$^2$ \cite{aid} and
c) $0.147<x<0.869$, $Q_{max}^2=4$~GeV$^2$ \cite{kar}.
We first consider the $p_T$ distribution $d\sigma/dp_T$ integrated over the 
rapidity interval $-1.5<y_{lab}<1$.
In Figs.~\ref{fig8}a--c, we confront the H1 tagged, H1 untagged and ZEUS
untagged data with our respective NLO predictions.
The renormalization and factorization scales are chosen to be
$\mu=M_f/2=\xi m_T$ with $\xi=1/2$ (dashed lines), $\xi=1$ (solid lines) and
$\xi=2$ (dash-dotted lines).
We observe that the scale dependence is small, approximately $\pm10\%$, 
indicating that corrections beyond NLO are likely to be negligible.
In all three cases, the agreement with the data is remarkably good, even at
small $p_T$, where we would not have expected it.
In our approach, the $p_T$ spectra come out somewhat larger than in the
all-massive theory \cite{sfr} which was used for comparison in
\cite{aid,kar} and therefore agree slightly better with the data.
In the low-$p_T$ range, this may be partly attributed to the omission of the
power terms in $m/p_T$ in our approach.

It is well known that the rapidity distribution $d^2\sigma/dy_{lab}\,dp_T$ at 
fixed $p_T$ usually allows for a more stringent test of the theory than
$d\sigma/dp_T$.
Unfortunately, the measured $y_{lab}$ distributions available so far all 
correspond to $d^2\sigma/dy_{lab}\,dp_T$ integrated over $p_T$ intervals
with rather small lower bounds, namely 2.5~GeV${}<p_T<{}$10~GeV in the case of
H1 \cite{aid} and 3~GeV${}<p_T<{}$12~GeV in the case of ZEUS \cite{kar}.
The corresponding H1 tagged, H1 untagged and ZEUS untagged data are compared 
with our NLO predictions for $\xi=1/2$, 1 and 2 in Figs.~\ref{fig9}a--c,
respectively.
The theoretical predictions for H1 tagged and H1 untagged are quite different.
The curves for the tagged case have their maxima at smaller $y_{lab}$, 
close to $y_{lab}=-1.2$, and exhibit a much stronger variation with $y_{lab}$
than the curves for the untagged case, which reach their maxima near $-0.5$.
This difference may be understood by observing that the soft end of the 
equivalent photon spectrum is eliminated in the tagged case by the cut
$x>0.28$.
The agreement between theory and data is worse than in the case of
$d\sigma/dp_T$.
On the other hand, the experimental errors are still rather large.
We observe that in all three cases there is perfect agreement in the central
region of the detector, at $y_{lab}=-0.25$, whereas in the backward direction
the theoretical predictions tend to slightly overshoot data, especially in the 
case of H1.
In the case of ZEUS, where the lower cut on $p_T$ is somewhat larger than for 
H1, our prediction agrees slightly better with the data than the result of
the all-massive calculation presented in Fig.~4 of \cite{kar}.
However, this observation has to be taken with a grain of salt, since the bulk
of the cross section is accumulated at the low-$p_T$ end, where our approach
is expected to be less reliable.

In the comparison of the data with our predictions, we have to keep in mind 
that the latter depend on the parameter $\epsilon$ and the starting scale 
$\mu_0$ of the Peterson FF.
We have chosen $\epsilon=0.06$ and $\mu_0=2m$.
In Fig.~\ref{fig7}, we have seen that varying $\epsilon$ by $\pm0.02$ changes 
the cross section by $\pm15\%$, which is comparable to the scale dependence.
We have verified by explicit calculation that the choice $\epsilon=0.06$ in 
conjunction with $\mu_0=2m$ yields a satisfactory global fit to the
$\langle x\rangle$ values measured in $e^+e^-$ experiments with CM energies
between 10 and 91.2~GeV.
We conclude that, at the present stage, the comparison between
theory and data is not jeopardized by the uncertainty in $\epsilon$.

\section{Summary and Conclusions}

In this work, we calculated the cross section of inclusive 
$D^{*\pm}$-meson production via small-$Q^2$ $ep$ scattering at HERA energies 
on the basis of a new massless-charm approach.
Specifically, the $\overline{\mbox{MS}}$ factorization of the collinear
singularities associated with massless charm quarks in the final state is
adjusted so as to match the corresponding massive-charm calculation.
This is implemented in a way similar to switching from the 
$\overline{\mbox{MS}}$ factorization scheme e.g.\ to the
deep-inelastic-scattering (DIS) scheme within the massless analysis.
The fragmentation of the quasi-massive charm quarks into the physical
$D^{*+}$ mesons at long distances is then described by the NLO-evolved
Peterson FF with a suitable choice of $\epsilon$ at the starting scale $\mu_0$.
Compared to the all-massive calculation, this has the advantage that the large
logarithms $\ln(p_T^2/m^2)$ are resummed, in particular through the appearance
of a much more sizeable resolved-photon contribution.
We believe that this massless-charm scheme is much more suitable in the
large-$p_T$ regime.
Furthermore, it offers us the opportunity to investigate the charm 
distribution inside the resolved photon (see Fig.~\ref{fig5}), which does not
enter the all-massive calculation.

We studied various refinements of the massless-charm approach w.r.t.\ the 
description of the $c\to D^{*+}$ fragmentation.
The most realistic fragmentation model, based on the Peterson FF evolved
through NLO evolution equations from the starting scale $\mu_0=2m$ up to the
characteristic scale of the process, was examined in great detail and used for
comparisons with recent data from the H1 and ZEUS Collaborations at HERA.
We found good agreement, in particular for the $p_T$ spectra as well as the
$y_{lab}$ spectra in the central region.
Surprisingly, even at rather low $p_T$, some of our predictions agree with the
data slightly better than those obtained within the all-massive scheme
(compare Figs.~\ref{fig8}c and \ref{fig9}c with Fig.~4 of \cite{kar}),
which may be an artifact of neglecting the non-singular power terms in $m/p_T$.
Since our approach is better justified theoretically for $p_T\gg m$,
we hope that HERA data on inclusive $D^{*\pm}$ photoproduction in the
large-$p_T$ region will soon be available with high statistics.
In particular, $y_{lab}$ spectra with large minimum-$p_T$ cuts would allow
for a more stringent test of our fragmentation model in particular and
the QCD-improved parton model in general, and increase our understanding of
the charm distribution inside the resolved photon.

If one determines $\epsilon$ from an NLO fit to $e^+e^-$ data, the result will
depend on whether scheme $E$ or scheme $F$ is adopted.
According to the factorization theorem, this difference is expected to be
compensated if one makes NLO predictions for different kinds of experiments as 
long as one works in the respective scheme.
By the same token, NLO predictions based on the all-massive scheme with
Peterson fragmentation \cite{sfr} suffer from an essentially uncontrolled
normalization as long as the $\epsilon$ parameter is not fitted within the
same scheme.

After the completion of this work, new sets of $D^{*\pm}$ FF's have been
extracted from fits to $e^+e^-$ data adopting scheme E \cite{bin} and a
variant of scheme F \cite{gre}.

\bigskip
\centerline{\bf ACKNOWLEDGMENTS}
\smallskip\noindent
One of us (GK) thanks the Theory Group of the Werner-Heisenberg-Institut for
the hospitality extended to him during a visit when this paper was finalized.
One of us (MS) thanks P. Nason for useful discussions about \cite{mel}.

\begin{appendix}

\section{Mellin transform of the Peterson fragmentation function}

The Mellin transform of a distribution function, $D(x)$, is defined as
\begin{equation}
\tilde{D} (N) = \int_0^1 dx x^{N-1} D(x),
\end{equation}
where $N$ is complex.
Taking $D(x)$ to be the Peterson FF of (\ref{eqpssz}), we find
\begin{eqnarray}
\tilde{D} (N) &\n=\n& \frac{2{\cal N}}{(N+1)(N+2)(N+3)} \left[
    \frac{ _2F_1 \left(2,3;N+4;x_1^{-1}\right)}{x_1^2 (x_1-x_2)^2}
+   \frac{ _2F_1 \left(2,3;N+4;x_2^{-1}\right)}{x_2^2 (x_1-x_2)^2}
\right. \nonumber \\
&\n\n& \left. \hspace{1cm}
+ 2 \frac{ _2F_1 \left(1,3;N+4;x_1^{-1}\right)}{x_1   (x_1-x_2)^3}
- 2 \frac{ _2F_1 \left(1,3;N+4;x_2^{-1}\right)}{x_2   (x_1-x_2)^3}
\right],
\end{eqnarray}
where $\cal N$ is the normalization factor in (\ref{eqpssz}),
\begin{equation}
_2F_1 (a,b;c;z) = \frac{\Gamma(c)}{\Gamma(b) \Gamma(c-b)} 
\int_0^1 dx~x^{b-1} (1-x)^{c-b-1} (1-xz)^{-a}
\end{equation}
is the hypergeometric function and
\begin{equation}
x_{1/2} = \frac{\epsilon}{2} \left( 1 \pm i \sqrt{\frac{4}{\epsilon} - 1}
\right).
\end{equation}
The following expansion is useful for numerical purposes:
\begin{eqnarray}
_2F_1 (a,a+m;c;z) &\n=\n&
\frac{\Gamma(c) (-z)^{-a-m}}{\Gamma(a+m) \Gamma(c-a)}
\sum_{n=0}^\infty \frac{(a)_{m+n} (1-c+a)_{m+n}}{n! (m+n)!} z^{-n}
\left[ \ln (-z) \right. \nonumber \\
&\n\n& \left.
+ \psi(1+m+n) + \psi(1+n) -\psi(a+m+n) - \psi(c-a-m-n)
\right] \nonumber \\
&\n\n&
+ (-z)^{-a} \frac{\Gamma(c)}{\Gamma(a+m)} \sum_{n=0}^{m-1}
\frac{\Gamma(m-n) (a)_n}{n! \Gamma(c-a-n)} z^{-n},
\end{eqnarray}
where
\begin{equation}
\psi (x) = \frac{\Gamma'(x)}{\Gamma(x)}
\end{equation}
is the digamma function and
\begin{equation}
(a)_n = \frac{\Gamma(a+n)}{\Gamma(a)}
\end{equation}
is Pochhammer's symbol.

\end{appendix}

\newpage

\vskip-6cm

\begin{figure}

\centerline{\bf FIGURE CAPTIONS}

\caption{\protect\label{fig1} NLO $p_T$ distribution $d\sigma/dp_T$, integrated 
over $-1.5<y_{lab}<1$, of inclusive $D^{*\pm}$ photoproduction in $ep$
scattering with untagged positrons as in the ZEUS experiment.
The branching ratio $B(c\to D^{*+})$ is taken to be unity.
The fragmentation models $A$--$F$ are described in the text.\hskip5cm}

\vskip-.2cm

\caption{\protect\label{fig2} NLO $y_{lab}$ distribution
$d^2\sigma/dy_{lab}\,dp_T$, at $p_T=7$~GeV, of inclusive $D^{*\pm}$
photoproduction in $ep$ scattering with untagged positrons as in the ZEUS
experiment.
The branching ratio $B(c\to D^{*+})$ is taken to be unity.
The fragmentation models $A$--$F$ are described in the text.\hskip5cm}

\vskip-.2cm

\caption{\protect\label{fig3} NLO $y_{lab}$ distribution
$d^2\sigma/dy_{lab}\,dp_T$, at $p_T=7$~GeV, of inclusive $D^{*\pm}$
photoproduction in $ep$ scattering with untagged positrons as in the ZEUS
experiment.
The NLO-evolved Peterson FF is used in connection with the massive-charm
factorization scheme (case $E$).
The branching ratio $B(c\to D^{*+})$ is taken to be unity.
The contributions due to direct (dashed line) and resolved (dash-dotted line)
photoproduction are shown together with their sum (solid line).\hskip5cm}

\vskip-.2cm

\caption{\protect\label{fig4} Influence of the proton PDF's.
The total result of Fig.~\ref{fig3} with the CTEQ4M set (solid line) is
compared with the corresponding calculations with the charm density inside the
proton switched off (dashed line) and with the MRS(G) set (dash-dotted
line).\hskip5cm}

\vskip-.2cm

\caption{\protect\label{fig5} Influence of the photon PDF's.
The total result of Fig.~\ref{fig3} with the GRV set (solid line) is compared
with the corresponding calculations with the charm density inside the photon
switched off (dashed line) and with the GS set (dash-dotted line).
For comparison, also the direct-photon contribution of Fig.~\ref{fig3} is 
shown (dotted line).\hskip5cm}

\vskip-.2cm

\caption{\protect\label{fig6} Dependence on $\xi$, where
$\mu=M_f/2=\xi m_T$.
The total result of Fig.~\ref{fig3} with $\xi=1$ (solid line) is compared with
the corresponding calculations with $\xi=1/2$ (dashed line) and 2 (dash-dotted
line) as well as the one with $\mu=M_f=m_T$ (dotted line).\hskip5cm}

\vskip-.2cm

\caption{\protect\label{fig7} Dependence on the parameters $\epsilon$ and 
$\mu_0$ of the Peterson FF.
The total result of Fig.~\ref{fig3} with $\epsilon=0.06$ and $\mu_0=2m$ (solid
line) is compared with the corresponding calculations with $\epsilon=0.04$
(dashed line), $\epsilon=0.08$ (dash-dotted line) and $\mu_0=m$ (dotted line).
\hskip5cm}

\vskip-.2cm

\caption{\protect\label{fig8} The $p_T$ distributions $d\sigma/dp_T$,
integrated over $-1.5<y_{lab}<1$, of inclusive $D^{*\pm}$ photoproduction in
$ep$ scattering with (a) tagged and (b) untagged positrons as measured by H1
and with (c) untagged positrons as measured by ZEUS are compared with the
corresponding NLO predictions with $\xi=1$ (solid lines), $\xi=1/2$ (dashed
lines) and $\xi=2$ (dash-dotted lines).\hskip5cm}

\vskip-.2cm

\caption{\protect\label{fig9} The $y_{lab}$ distributions $d\sigma/dy_{lab}$,
integrated over $p_T$, of inclusive $D^{*\pm}$ photoproduction in $ep$
scattering with (a) tagged and (b) untagged positrons as measured by H1
(2.5~GeV${}<p_T<{}$10~GeV) and with (c) untagged positrons as measured by ZEUS
(3~GeV${}<p_T<{}$12~GeV) are compared with the corresponding NLO predictions
with $\xi=1$ (solid lines), $\xi=1/2$ (dashed lines) and $\xi=2$ (dash-dotted
lines).\hskip5cm}

\end{figure}

\newpage
\begin{figure}[ht]
\epsfig{figure=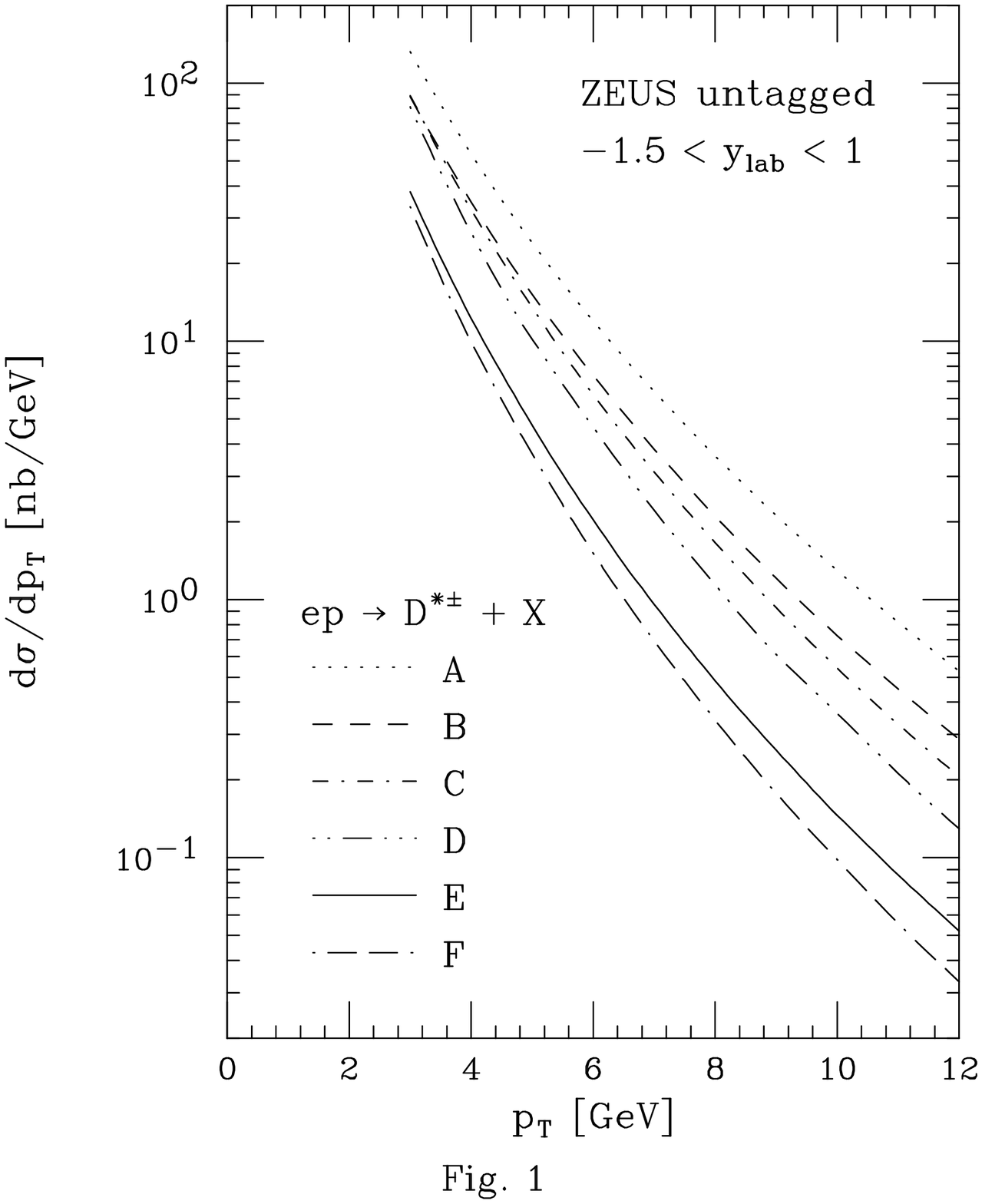,width=\textwidth}
\end{figure}
\newpage
\begin{figure}[ht]
\epsfig{figure=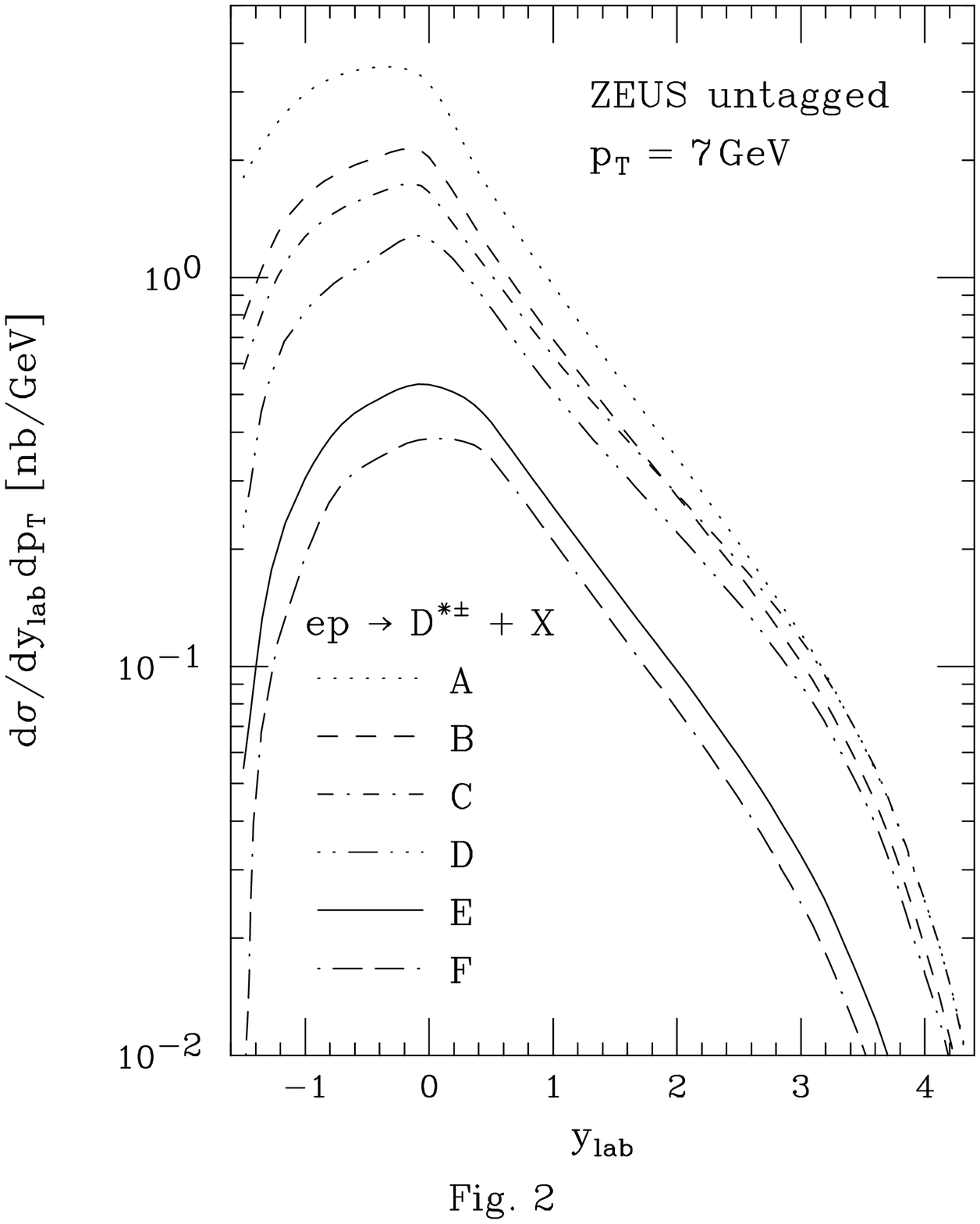,width=\textwidth}
\end{figure}
\newpage
\begin{figure}[ht]
\epsfig{figure=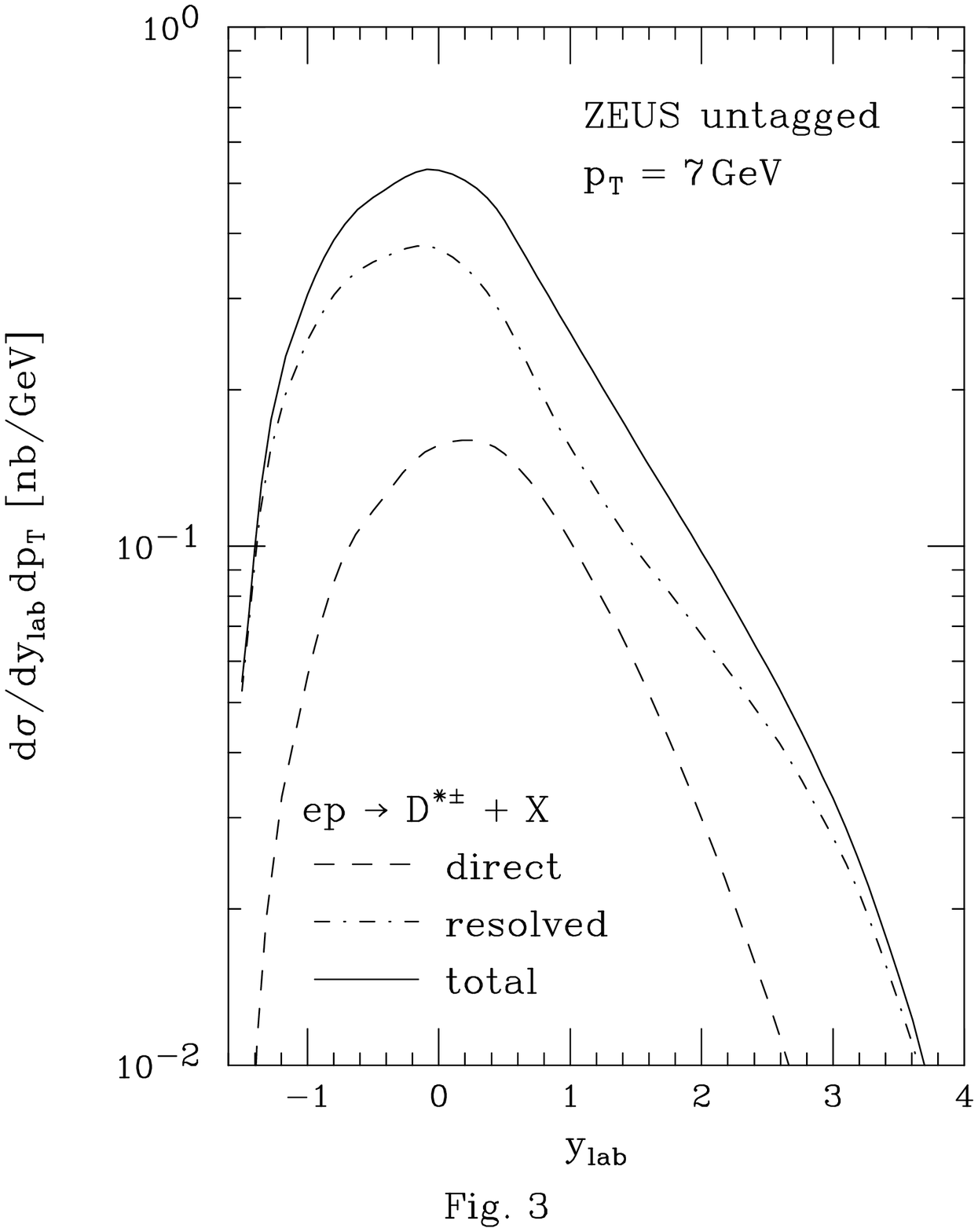,width=\textwidth}
\end{figure}
\newpage
\begin{figure}[ht]
\epsfig{figure=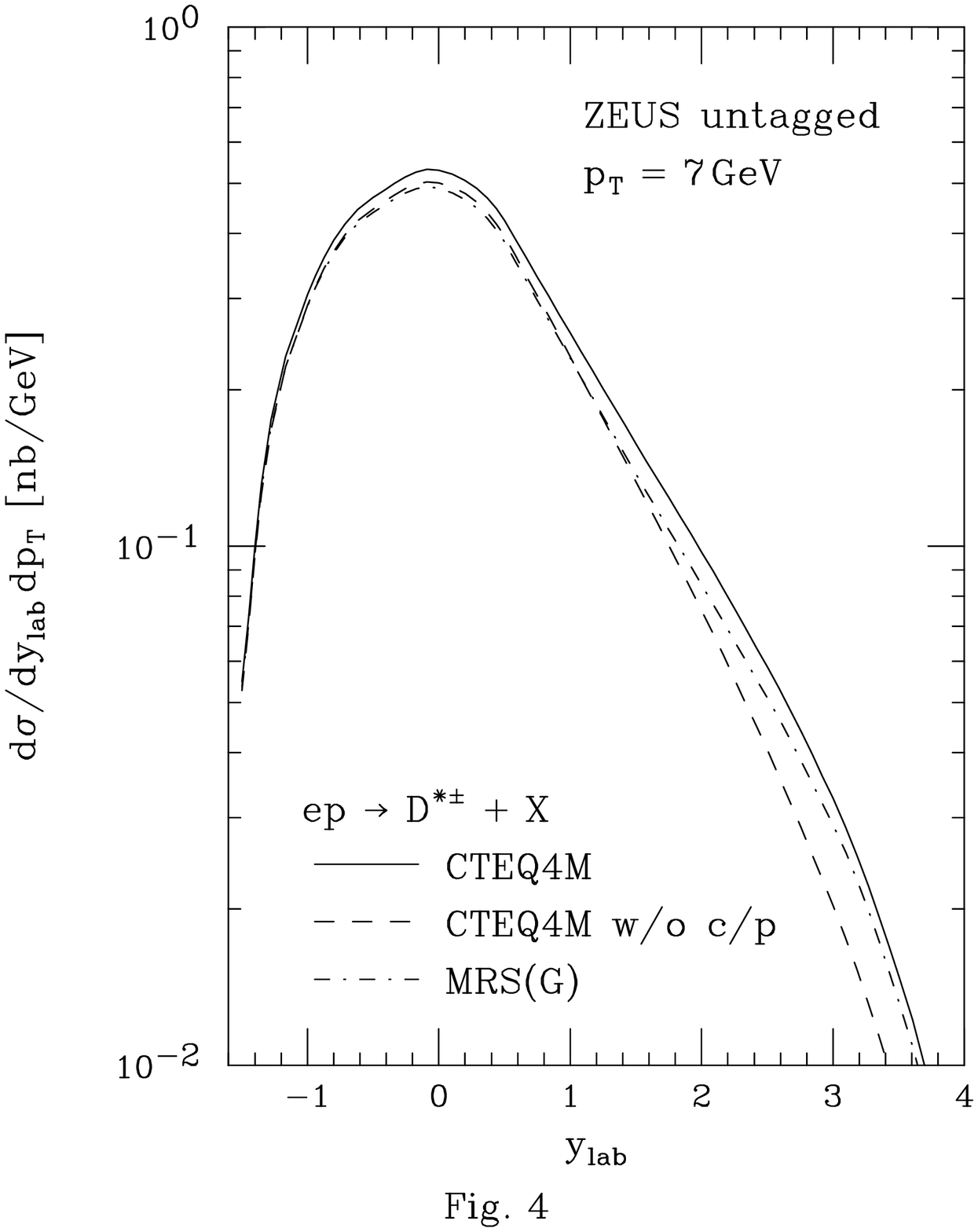,width=\textwidth}
\end{figure}
\newpage
\begin{figure}[ht]
\epsfig{figure=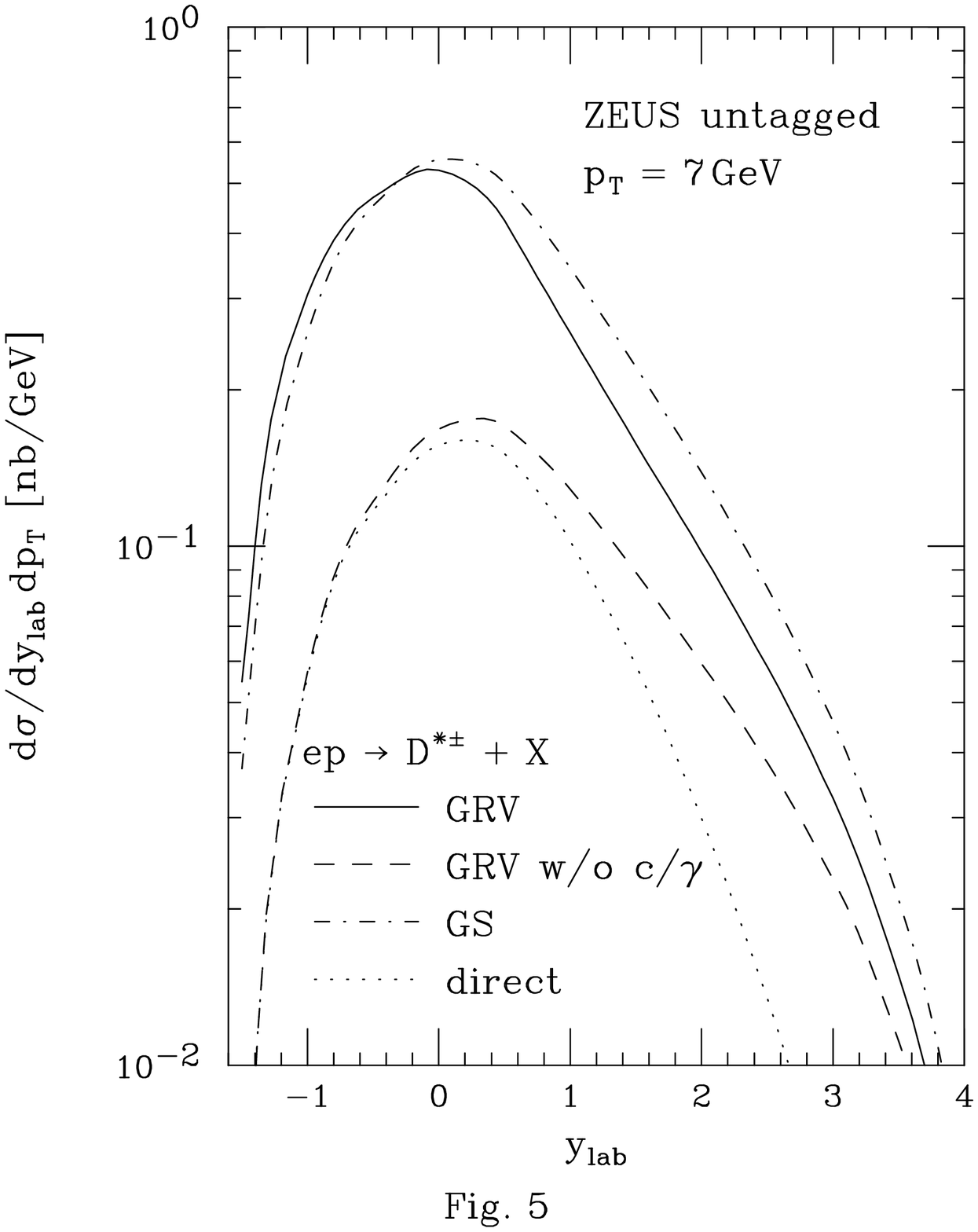,width=\textwidth}
\end{figure}
\newpage
\begin{figure}[ht]
\epsfig{figure=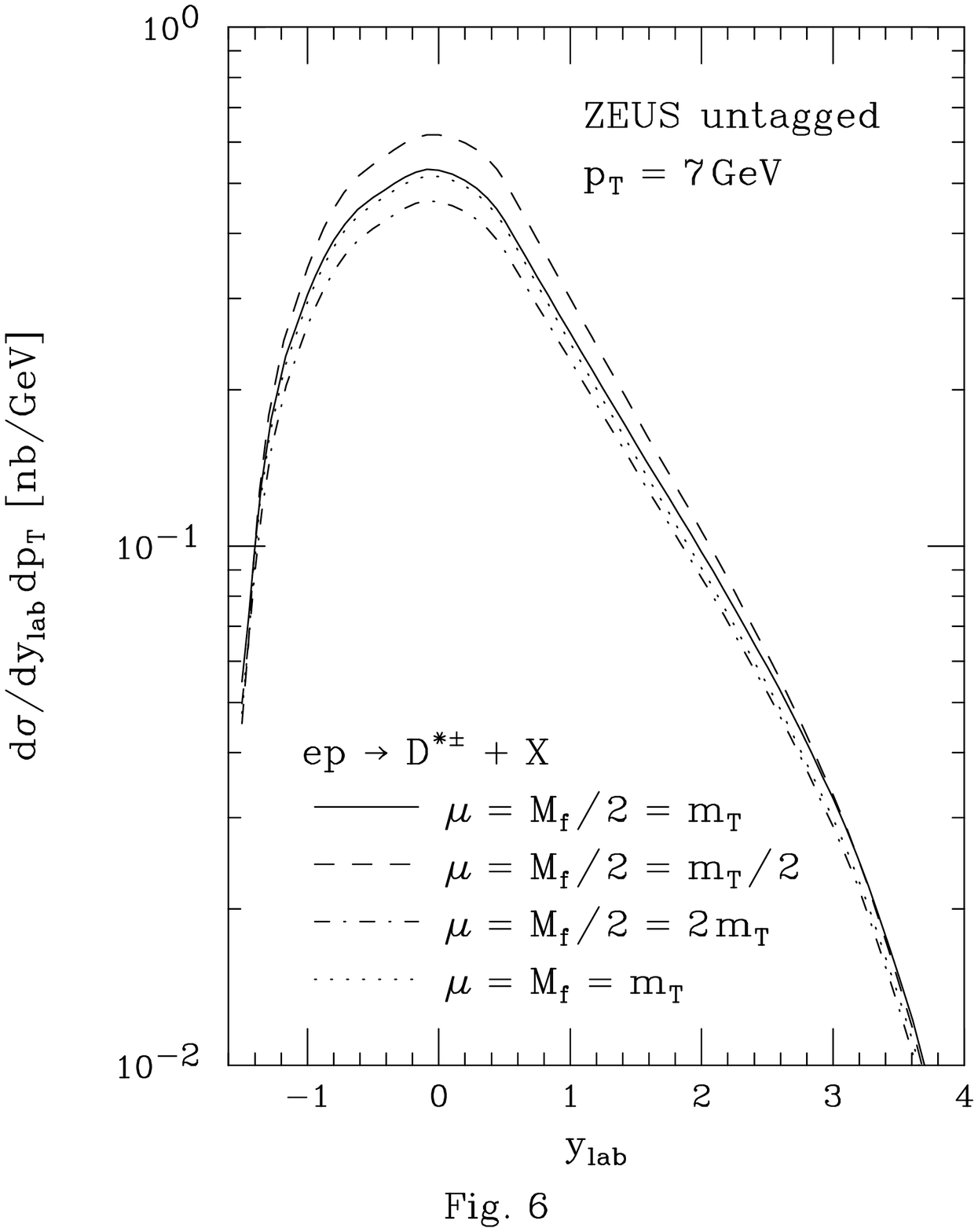,width=\textwidth}
\end{figure}
\newpage
\begin{figure}[ht]
\epsfig{figure=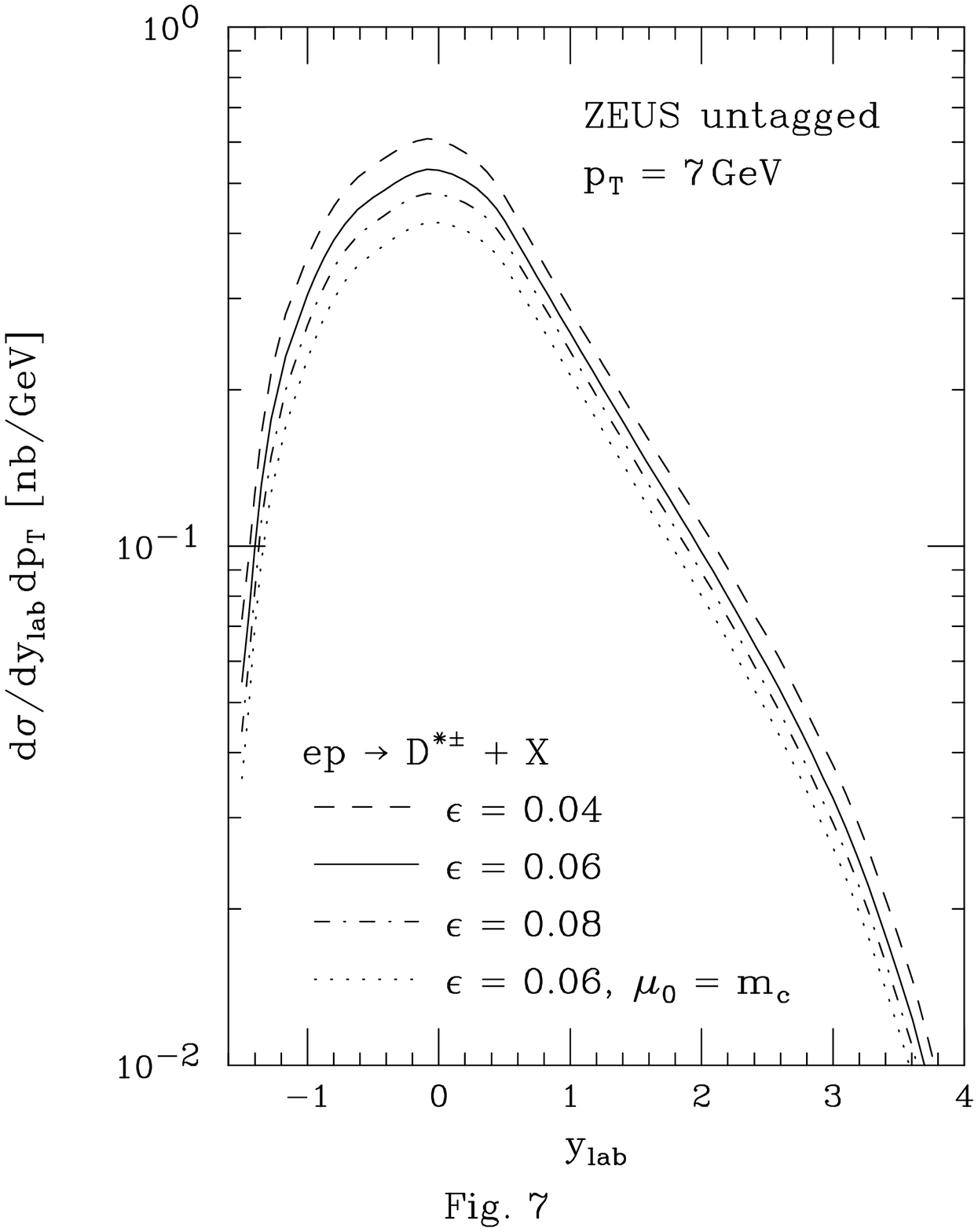,width=\textwidth}
\end{figure}
\newpage
\begin{figure}[ht]
\epsfig{figure=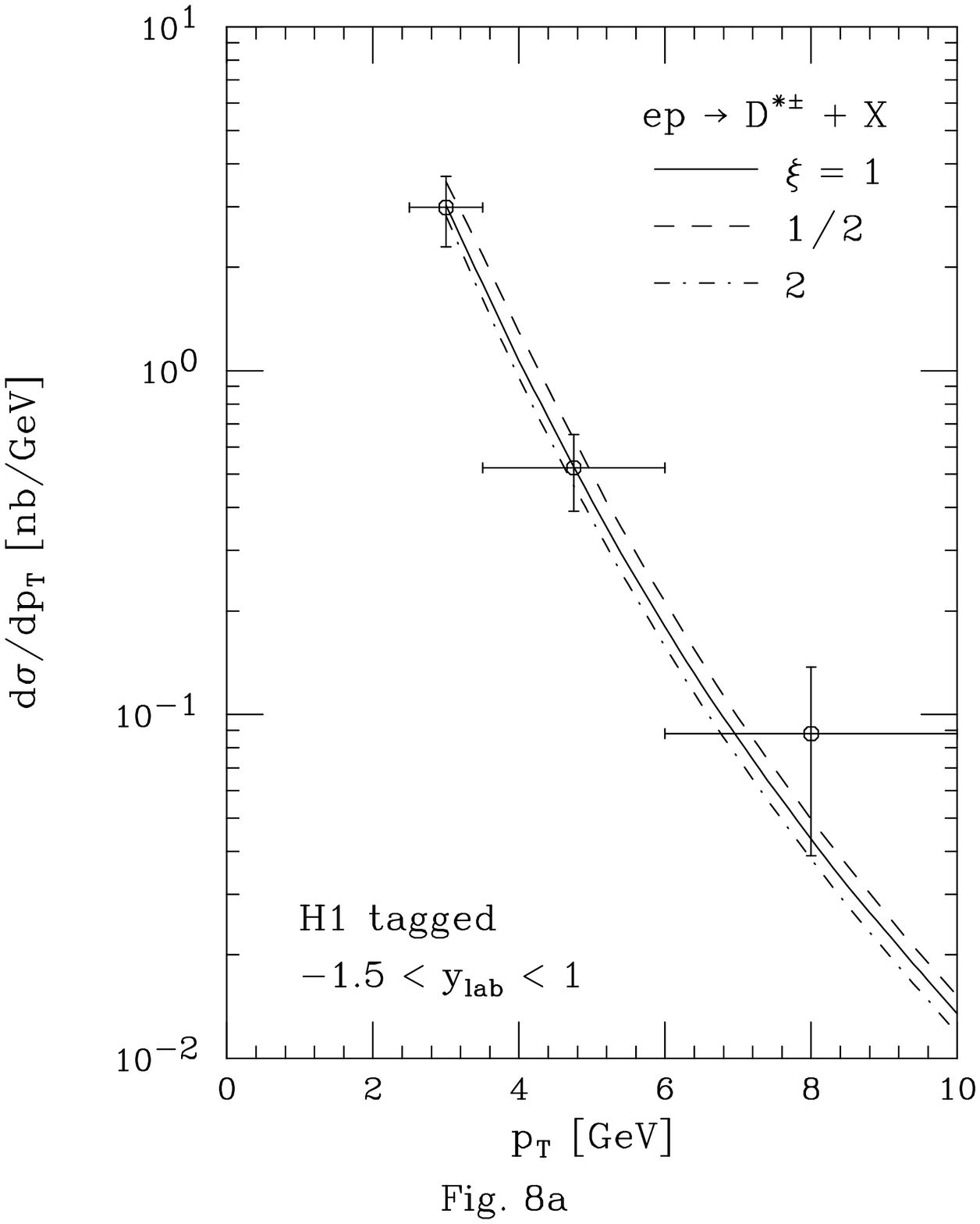,width=\textwidth}
\end{figure}
\newpage
\begin{figure}[ht]
\epsfig{figure=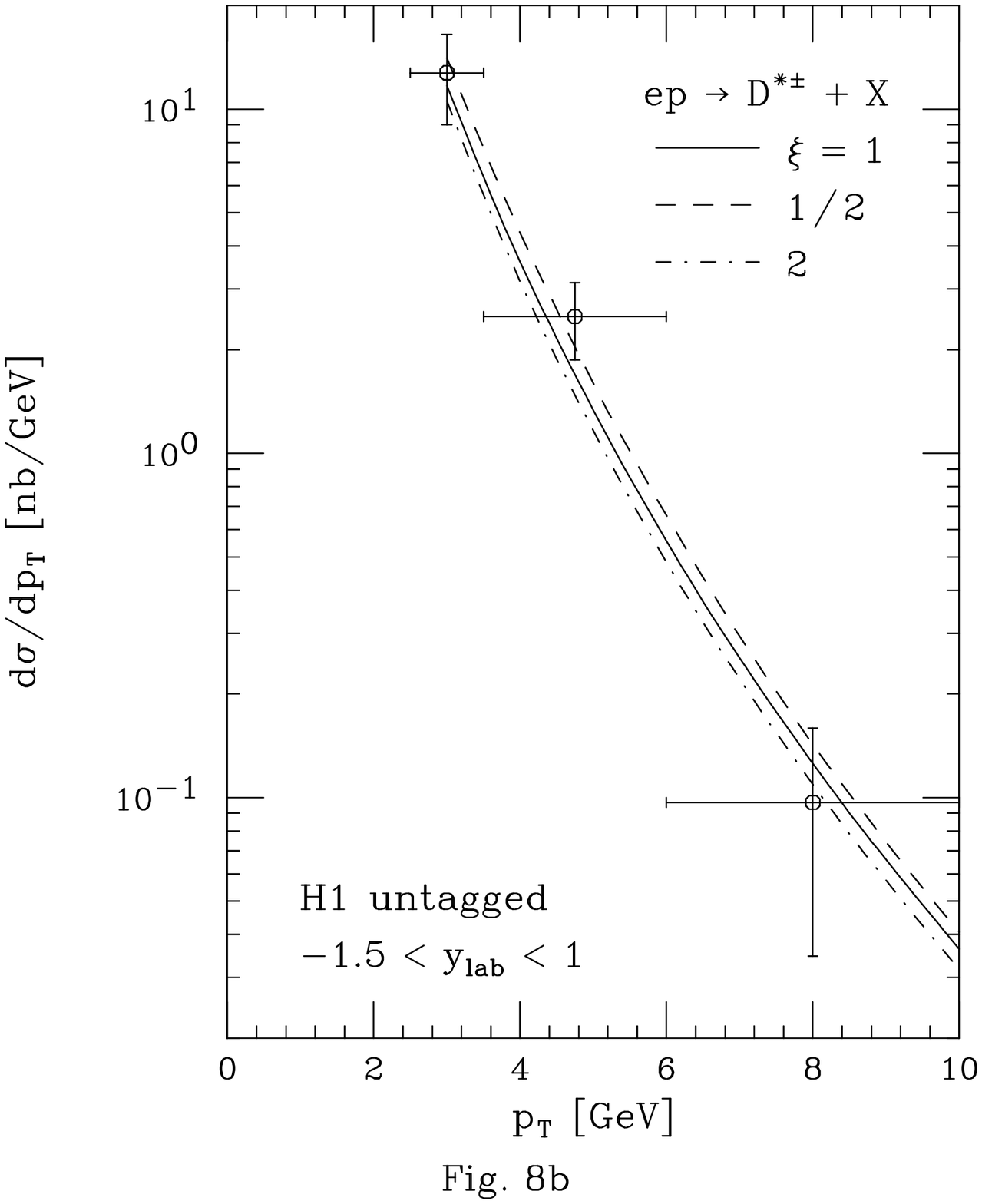,width=\textwidth}
\end{figure}
\newpage
\begin{figure}[ht]
\epsfig{figure=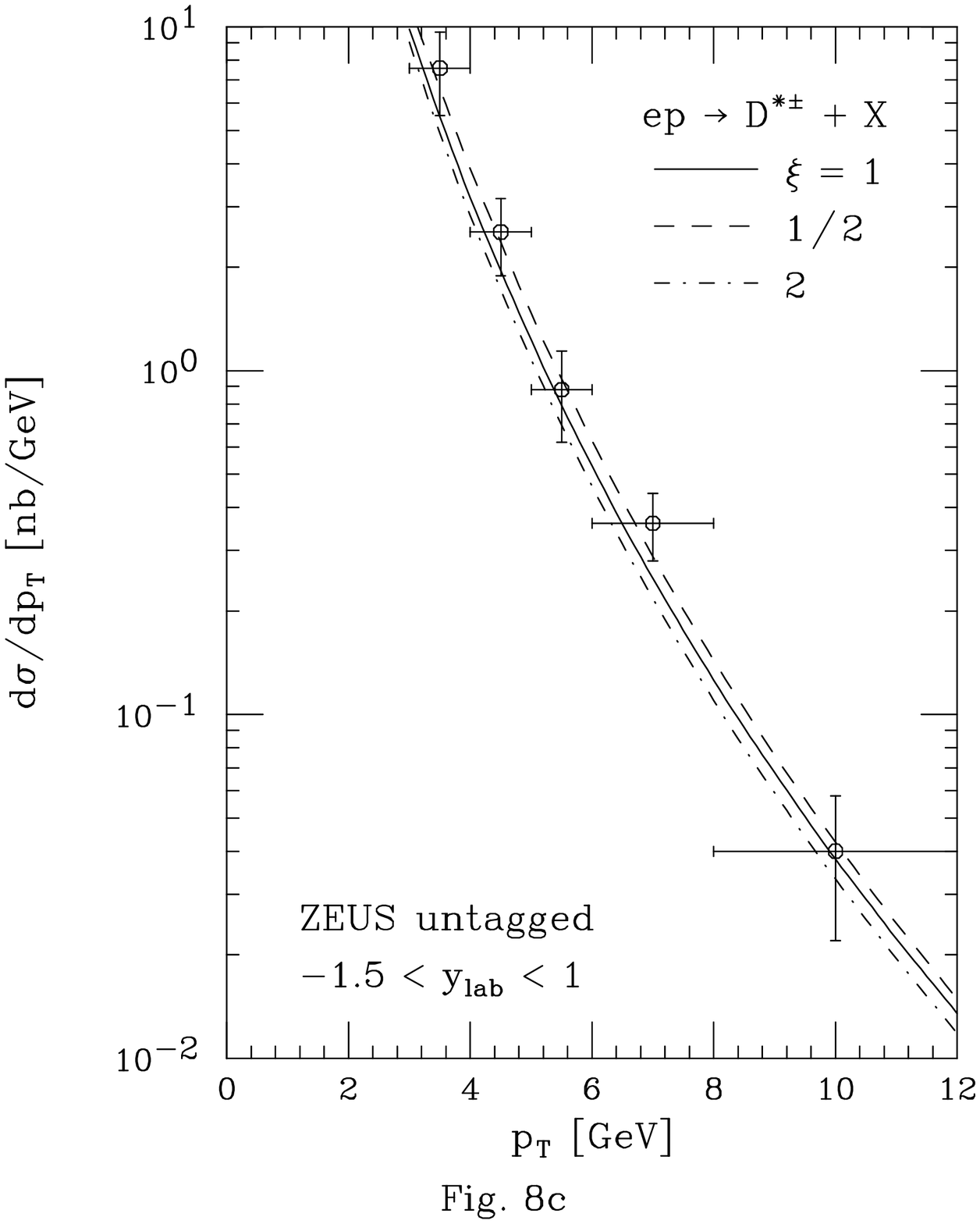,width=\textwidth}
\end{figure}
\newpage
\begin{figure}[ht]
\epsfig{figure=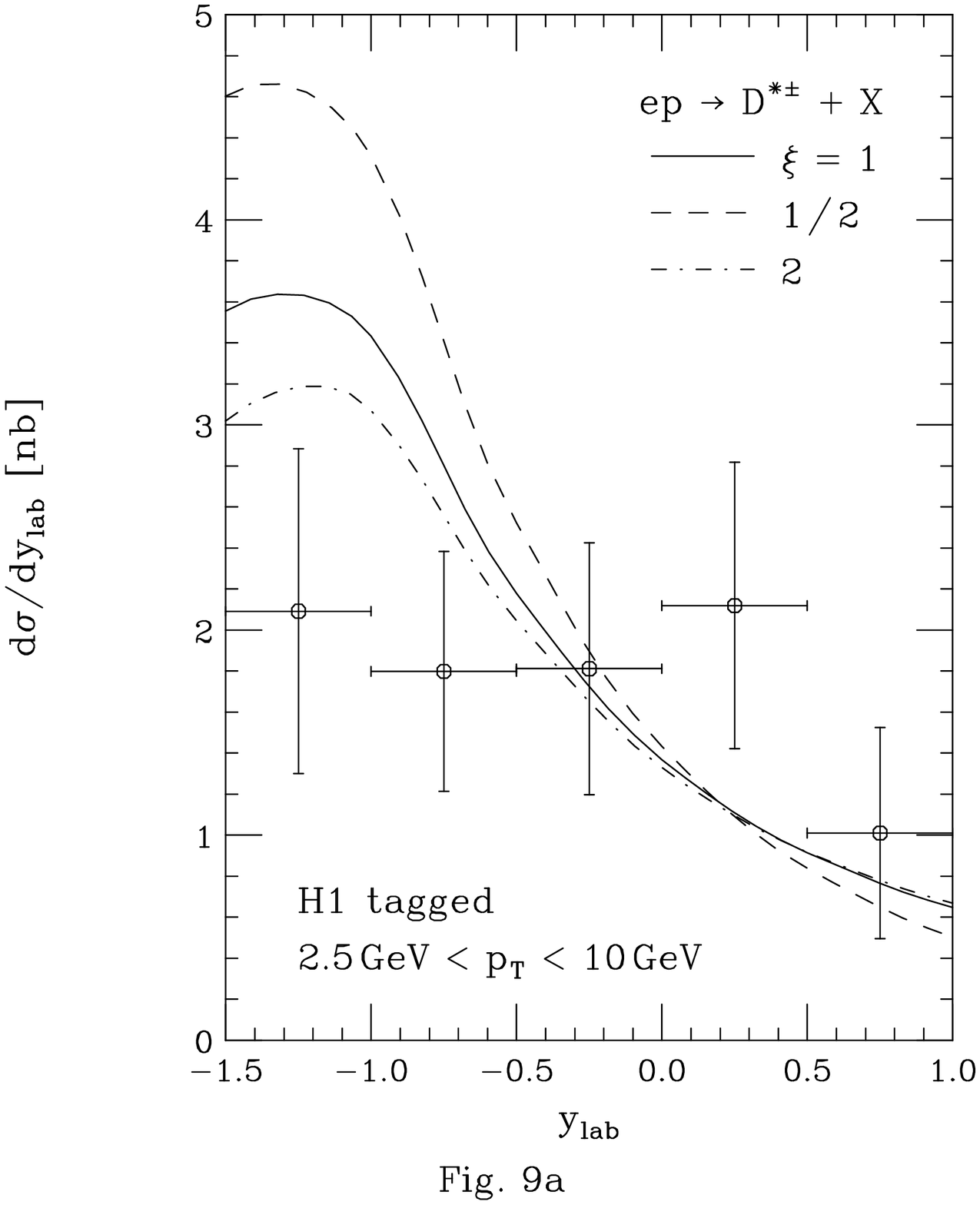,width=\textwidth}
\end{figure}
\newpage
\begin{figure}[ht]
\epsfig{figure=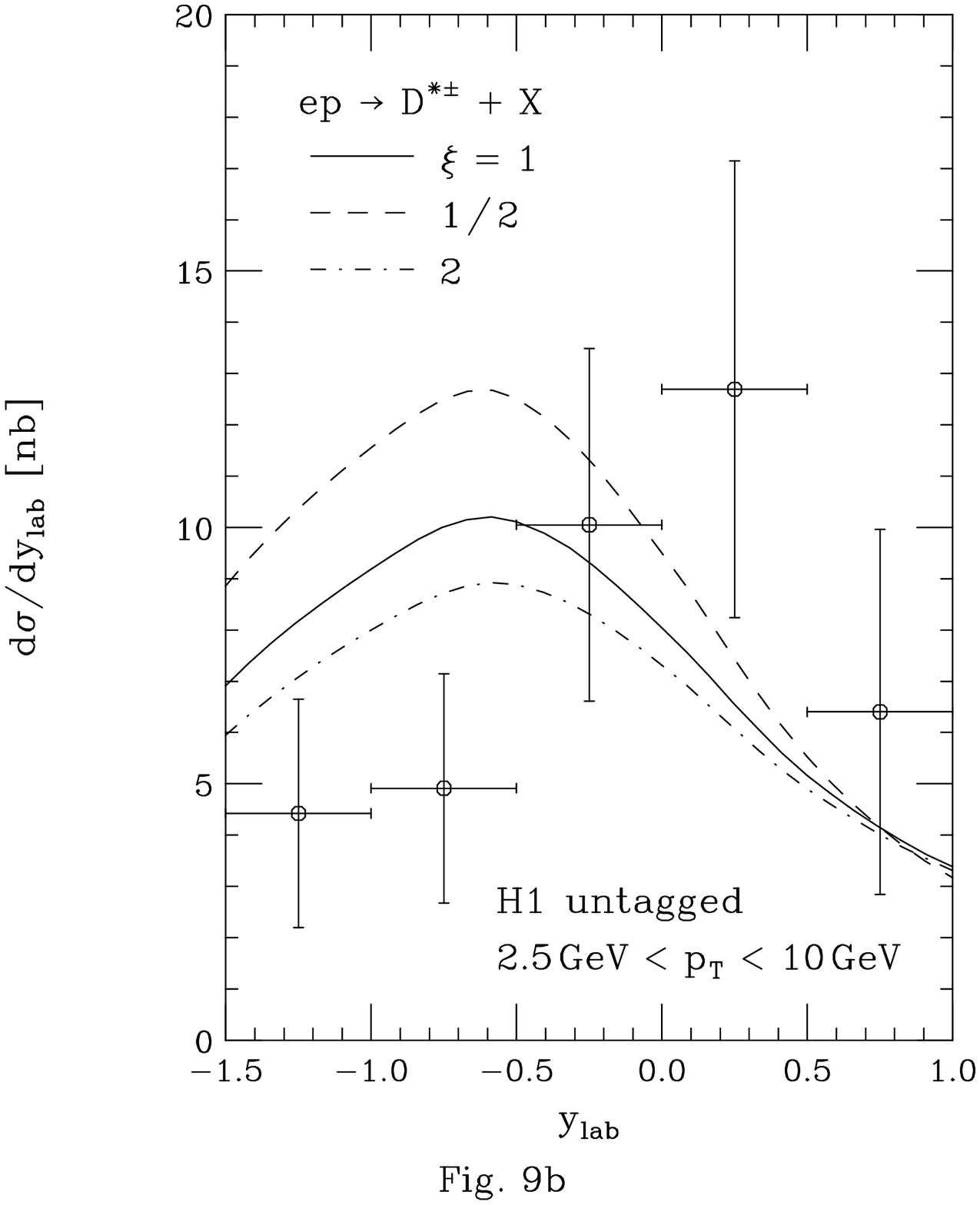,width=\textwidth}
\end{figure}
\newpage
\begin{figure}[ht]
\epsfig{figure=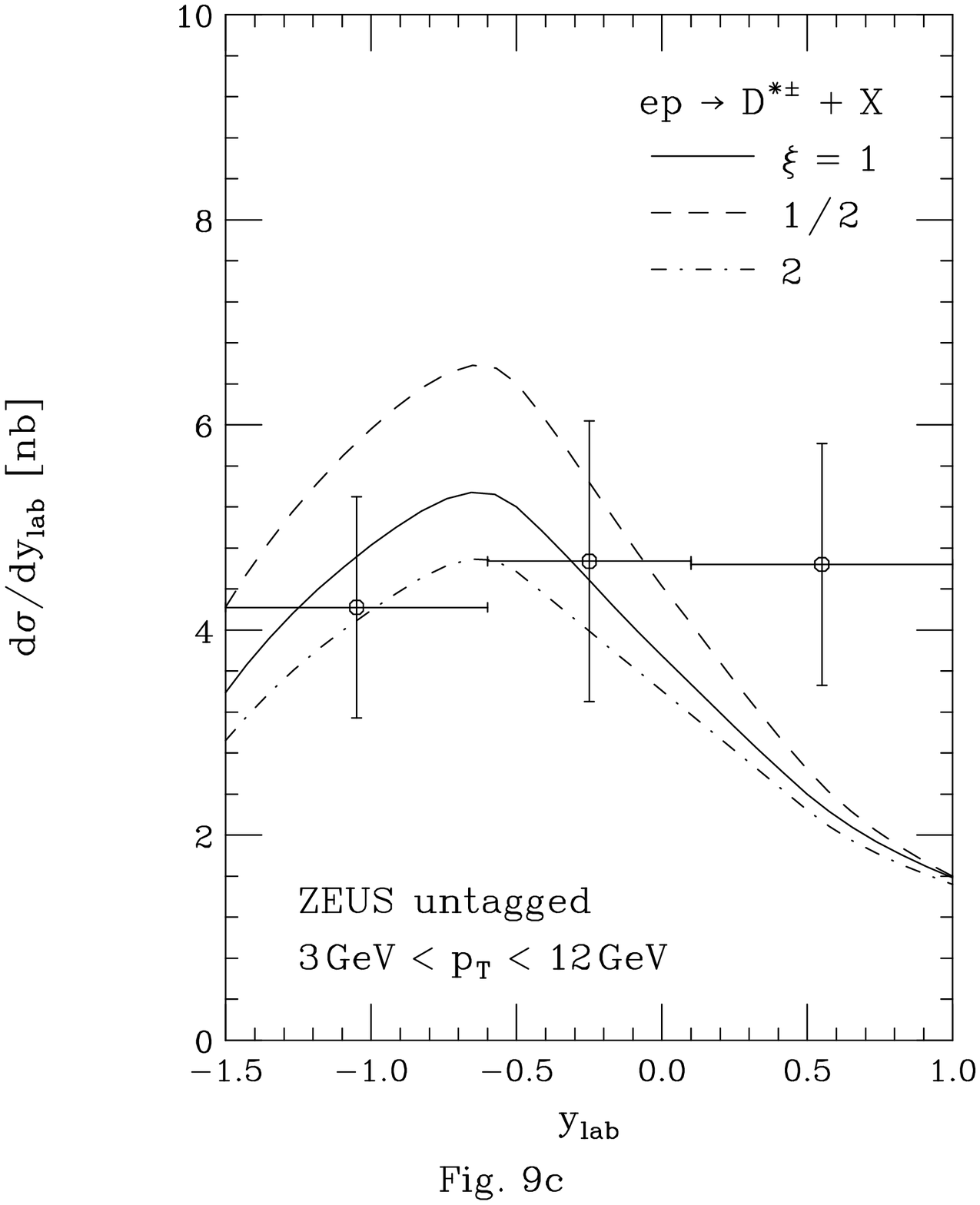,width=\textwidth}
\end{figure}

\end{document}